# Authorization Architectures for Tool-Using AI Agents

Rakesh Kumar Surapani[1], Pradeep Kumar Dolabehera Kakitapelli[2], Arun Morampudi[3], Praveena Padi[4]*

[1] Westcliff University, Irvine, California, USA

[2] University of the Cumberlands, Williamsburg, Kentucky, USA

[3] Delta Air Lines, Inc., Atlanta, Georgia, USA

[4] Georgia Institute of Technology, Atlanta, Georgia, USA

Corresponding author: Praveena Padi (ppadi7@gatech.edu)

## Abstract

Tool-using artificial intelligence (AI) agents, systems that autonomously invoke application programming interfaces (APIs), databases, browsers, and inter-agent protocols such as the Model Context Protocol (MCP), are becoming production infrastructure. However, the security model governing when an agent is authorized to act on a human's behalf remains underdeveloped. Trustworthy human–AI systems require that every consequential agent action be traceable to a human principal, bounded by what that human actually delegated, and contestable afterwards; few documented deployments satisfy all three properties reliably and end to end. Existing studies address fragments in isolation: credential management for non-human identities, classical access control models, prompt injection, and audit trails. Less attention has been given to the authorization decision point, the moment a tool invocation occurs, and to the mechanisms that make that decision correct, enforceable, and accountable. This review introduces a principal hierarchy, human user, operator/deployer, orchestrator agent, sub-agent, and tool endpoint, as an organizing framework, and examines five interdependent layers: agent identity and credential lifecycle; delegation and scope propagation across multi-hop chains; runtime enforcement and just-in-time authorization at policy enforcement points (PEPs); prompt injection as an authorization bypass that breaks the principal hierarchy; and auditability, provenance, and non-repudiation. Drawing on a structured narrative review of 89 primary sources screened from approximately 180 candidates (2023–2026), we propose seven structural requirements, derive a four-layer reference architecture, apply the requirements to three deployable reference configurations, and identify runtime enforcement and aggregation bounds as the principal unresolved gaps.

## 1. Introduction

Large language models (LLMs) are increasingly designed not as text generators but as tool-using agents that plan, call external functions, read or modify external state, and perform multiple tasks. Early work has framed language models as components of modular architectures [1]. Subsequent systems made tool use explicit, trained models to decide when to call application programming interfaces (APIs) [2], and interleaved reasoning with actions such as searching or interacting with the environment [3]. More recent frameworks have extended this to persistent and multi-agent workflows [4,5]. Together, these mark an architectural shift from isolated prompt-response systems to agents that affect external computational environments, such as the Internet.

The tool layer is central to this process. An empirical study of public Model Context Protocol (MCP) repositories identified 177,436 agent tools created between November 2024 and February 2026, with an increasing share of action-oriented tools [6]. Surveys of agent communication protocols similarly find the ecosystem moving toward standardized interfaces for tool discovery and inter-agent coordination while remaining immature in terms of security, identity, and interoperability [7]. Therefore, we do not treat agent tooling as an implementation detail but as an operational boundary, where an LLM-generated plan becomes an externally visible action.

This boundary creates a critical authorization gap in the system. Classical security work emphasizes least privilege, complete mediation, and explicit protection boundaries [8], and access control models formalize authorization as a property of specific operations rather than as a program [9]. Tool-using agents strain these assumptions because the actor that chooses an action, the user who requested the task, the runtime that holds credentials, and the service that receives the request may be distinct entities. This separation reintroduces the confused deputy problem, in which an intermediary acts with authority that the original requester may not legitimately possess for a specific operation [10].

Recent research has shown that this gap is not only theoretical in nature. Goel [11] identified ambient authority leakage, overprivileged tools, and capability-intent mismatch as recurring risks in privileged execution environments. Bühler et al. [12] showed that MCP servers frequently execute with broad host access and propose declarative access control policies. Stanley et al. [13] frame personal-agent privacy as an enforcement problem that requires permission tracking across execution steps. Empirical work has shown that tool-using agents can perform complex offensive actions when given suitable capabilities [14], and prompt injection studies have shown that indirect or black-box injection can manipulate whether LLM-integrated applications invoke APIs or disclose data [15,16]. Therefore, agent safety cannot be reduced to model alignment; explicit authorization controls are required for the actions agents take.

We propose the tool invocation moment as the fundamental unit of analysis. Tool invocation occurs whenever an agent attempts to call an external API, executes code, retrieves private data, updates a file, sends a message, initiates a transaction, or causes an effect beyond token generation. This moment is security-critical because identity, intent, permission, provenance, and policy must converge now. Treating invocation as the unit of analysis aligns with complete mediation: every attempted access can be checked at the point of use rather than implicitly allowed by a broad credential granted at initialization [8].

To structure the analysis, we introduce a principal hierarchy for agentic systems. In access control theory, a principal is an entity whose authority can be represented, constrained, and audited [9]. In agentic workflows, authority is layered: a human initiates a task, an agent interprets and decomposes it, sub-agents, tools, or runtimes execute the intermediate steps, and services receive concrete requests. If user *U* delegates to agent *A* and *A* invokes tool *T*, the security question is not whether *A* holds a credential but whether *T*'s specific action is authorized by the correct principal within the intended scope in the current context. This framing allows us to reason about delegation boundaries, scope propagation, and accountability without treating the agent as a fully autonomous actor or a passive extension of the user.

**Scope and method.** This review focuses on the authorization and security of the interface between artificial intelligence (AI) agents and external tools. The reference set prioritizes peer-reviewed papers, conference and workshop papers, and research preprints. The Internet Engineering Task Force (IETF) standards, World Wide Web Consortium (W3C) specifications, and regulatory instruments are cited as primary evidence for protocol adoption or compliance. Industry and government reports are cited only for empirical statistics where no peer-reviewed alternative exists and are labeled as such in the References list. Vendor marketing was excluded from the study. The coverage runs through mid-2026.

The remainder of this paper is organized as follows. Section 2 provides the background and defines the scope, principal hierarchy, terminology, methods, and threat models of this study. Section 3 examines agent identity as a non-human identity lifecycle issue. Section 4 addresses delegation and scope propagation, including the confused deputy problem (CDP). Section 5 surveys the runtime enforcement at the tool-invocation layer. Section 6 reframes prompt injection as an authorization bypass mechanism. Section 7 covers auditability, provenance and non-repudiation. Section 8 synthesizes a unified authorization architecture, performs a requirements-versus-mechanisms gap analysis, applies the requirements to three deployable reference configurations, and states their consequences for human oversight. Section 9 discusses cross-cutting open challenges and future directions. Section 10 concludes.

**Contributions.** The underlying security principles (complete mediation [8], least privilege, and confused-deputy prevention [10]) are well established; no individual principle here is new. The contribution of this survey is its *integrative application* to the agentic context, revealing domain-specific failure modes that arise from the combination of multi-principal probabilistic planners, dynamic tool discovery, semantic delegation gaps, and untrusted-content-influenced decision-making. To the best of our knowledge, existing mechanisms do not address this combination, although each principle has been addressed in isolation. Specifically, (C1) we propose a *principal hierarchy* as a structural framework that models humans, operators, orchestrators, sub-agents, and tool endpoints as layered principals, providing a common vocabulary for reasoning about multi-hop authorization failures. (C2) We identify the *tool invocation moment* as the fundamental unit of analysis, an instantiation of complete mediation for semantically mediated, multi-principal tool selection, arguing that authorization should be decided per action rather than through broad, up-front trust. (C3) We reframe *prompt injections* as an authorization-bypass mechanism (a confused-deputy variant in which untrusted Layer-4 content borrows Layer-0 authority) rather than purely an adversarial ML problem and derive architectural consequences for the enforcement-layer design. (C4) We formulate seven structural requirements (R1–R7) for robust agent authorization, derived from classical principles but scoped to the specific failure modes that the agentic context reveals; the novelty lies in the scoping and in identifying which requirements remain unsatisfied (particularly R4, aggregation bounds), not in the principles. (C5) We map existing mechanisms from access control theory, capability design, identity and access management (IAM), privacy-preserving runtimes, and agent security frameworks onto these requirements, exposing the areas where the coverage is strong, partial, or absent, and we record the evidence behind each judgement in Appendix A. (C6) We apply the requirements to three stacks that organizations can deploy today, trace one denied invocation end to end from human instruction through enforcement to audit record, show how the outcome changes with the placement of the enforcement point, and derive a pre-deployment checklist (Sections 8.5 and 9.7). Throughout, the organizing concern is human authority: the point of an authorization architecture is that a person can delegate a bounded task, see what was done with that authority, intervene while the workflow is running, and contest the result afterwards. Together, these provide a research-grounded foundation for designing agent architectures whose external actions are attributable, bounded, enforceable, and contestable.

## 2. Background and Scope

This section describes the terminology, architectural assumptions, review method, and threat model used in this study. Its purpose is narrow: rather than surveying all agentic AI, it defines the authorization-relevant properties of tool-using agents and the boundaries within which identity, delegation, enforcement, injection resistance, and auditability are analyzed.

### 2.1 What makes tool-using agents architecturally distinct

Software agent theory predates LLMs, emphasizing autonomy, reactivity, proactiveness, and social ability [17]. These properties remain necessary but no longer suffice to characterize contemporary tool-using agents. Earlier systems operated over explicitly modeled environments and predefined action spaces; modern LLM agents translate natural language goals into plans, select tools from schema descriptions, call external services, process returned data, and revise actions over long horizons [18–20]. The consequential shift is not that an LLM replaces a rule-based planner, but that it becomes a runtime mediator among human intent, operator policy, tool descriptions, credentials, memory, external data, and downstream effects [2,3,21].

This changes the security issues. In conventional applications, authorization is bound to an authenticated user, a route, an object, or a transaction. In a tool-using agent, the relevant action may be separated from the original request by planning steps, retrieving documents, intermediate tool outputs, sub-agent delegation, and model-generated arguments. The system must therefore ask not only "which credential is used?" but "whose authority is exercised, for what task, through which delegation chain, against which endpoint, with what constraints, and with what audit evidence?" This makes the tool invocation moment a natural unit for analysis.

### 2.2 The tool invocation stack and definition

Tool invocation is a stack of interfaces, schemas, protocols, runtimes, and policy layers through which an agent converts the intended operation into an external function. At the lowest level, tools are ordinary capabilities such as HTTP APIs, database queries, file operations, command-line programs, browser automation, code sandboxes, and enterprise workflow actions. LLM-facing systems represent these as callable tools: a model is shown tool definitions, emits a structured call, the host executes it, and the returned observation may influence subsequent reasoning [2,18,21]. Modern agents increasingly discover, compose, and invoke heterogeneous capabilities across workflows instead of operating in a single fixed action space [19,21]. This compositionality reduces integration friction but intensifies the authorization problem because agents may combine tools across organizational, data, and trust boundaries.

In this survey, a *tool invocation* is defined as any model-mediated request that causes a system outside the model inference process to read, write, compute, transmit, transform, retrieve, or act on information. This includes low-risk reads and high-consequence writes, both of which are authorization-relevant, even though their consequences differ. The security question is not only whether an endpoint is reachable but also whether the invocation is permitted under the authority, task scope, data constraints, and temporal conditions associated with the principal chain.

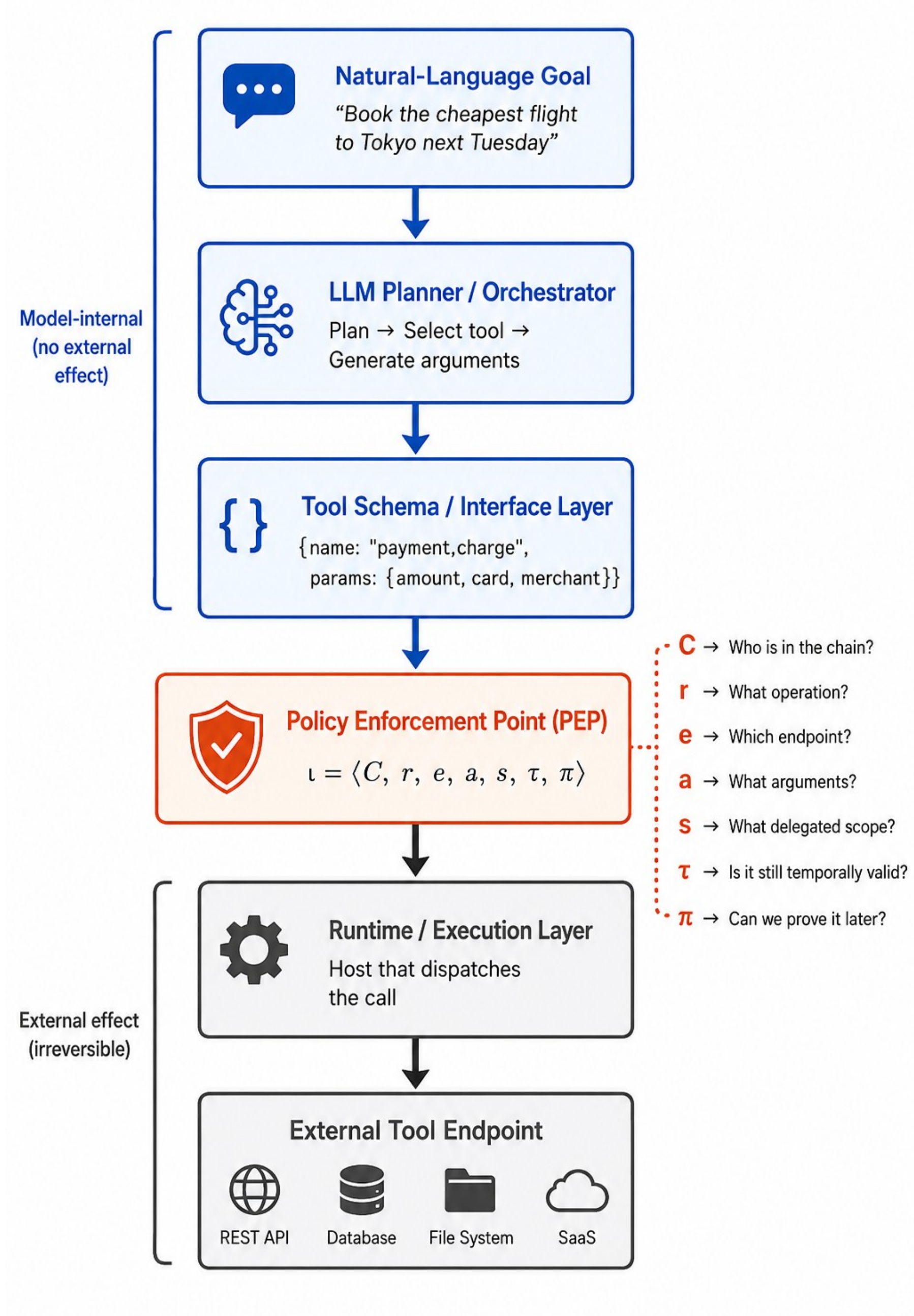

**Fig. 1** The tool invocation stack, from natural-language goal to external endpoint, with the policy enforcement point (PEP) as the authorization boundary

### 2.3 The principal hierarchy model

The organizing model has a principal hierarchy. A *principal* is any entity whose authority, identity, policy obligations, or accountability are relevant for invoking a tool. We define five layers:

**Layer 0: Human users or data subjects.** The human principal whose request, data, consent, or role initiates or constrains the workflow. This layer is the source of legitimate authority: every downstream action should be traceable to, and bounded by, what a Layer-0 principal delegated, and should remain contestable by that principal after the fact.

**Layer 1: Operator or Deployer.** The organization that configures the agent selects models and tools, sets policies, provisions credentials, and assumes operational responsibility. A user may request an action that operator policy prohibits; an operator may expose tools that a given user is not authorized to use.

**Layer 2: Orchestrator agent.** The primary LLM planner interprets and decomposes the task, selects tools, invokes subagents, and decides when to act. It is not an inherently trusted decision-maker; its proposed actions must be evaluated based on their identity, delegation, policy, and provenance.

**Layer 3: Sub-agents.** The orchestrator spawns specialized workers, planners, retrievers, and verifiers. They may introduce new trust boundaries, hidden prompts, separate memories, independent credentials, and opportunities for authority to drift.

**Layer 4: Tool Endpoints.** APIs, databases, file systems, Software as a Service (SaaS) applications, sandboxes, browsers, and registries that execute terminal operations. Endpoints are both targets of authorization and possible sources of untrusted content; an email inbox is a read target that also supplies adversarial instructions in message bodies.

**Two graphs, one decision.** The five layers describe *authority*, not network topology or dataflow, and conflating the two obscures the analysis. It is therefore useful to keep two distinct structures in view. The *principal and delegation graph* records who holds authority and who passed it to whom: Layer 0 → Layer 1 → Layer 2 → Layer 3 → service principals. It is directed and, in well-formed systems, monotonically attenuating; in multi-agent deployments it need not be a simple chain, as peer-to-peer, branching, and merging delegation all occur, so "hierarchy" describes the ordering of authority rather than a fixed five-node path. The *information and provenance graph* records where content came from and where it can flow: system instructions, user instructions, retrieved documents, memory entries, tool outputs, and terminal sinks. Layer 4 appears in both graphs, which is precisely why it is the locus of the prompt-injection problem (Section 6): a tool endpoint is simultaneously a target of authorization in the first graph and an untrusted source in the second. The authorization decision at the invocation boundary is the point at which the two graphs are evaluated together: the delegation graph determines what authority is available, and the provenance graph determines whether the content that motivated the proposed action is entitled to invoke it. Throughout the review, statements about attenuation and accountability refer to the delegation graph; statements about taint, aggregation, and injection refer to the provenance graph. We represent an invocation as where  is the ordered principal chain,  is the requested operation,  is the target endpoint or object,  is the argument,  is the delegated scope,  is the temporal/session context, and  is the provenance evidence available for audit. The notation is a common vocabulary for comparing identity, delegation, enforcement, and audit mechanisms across the survey, not formal semantics.

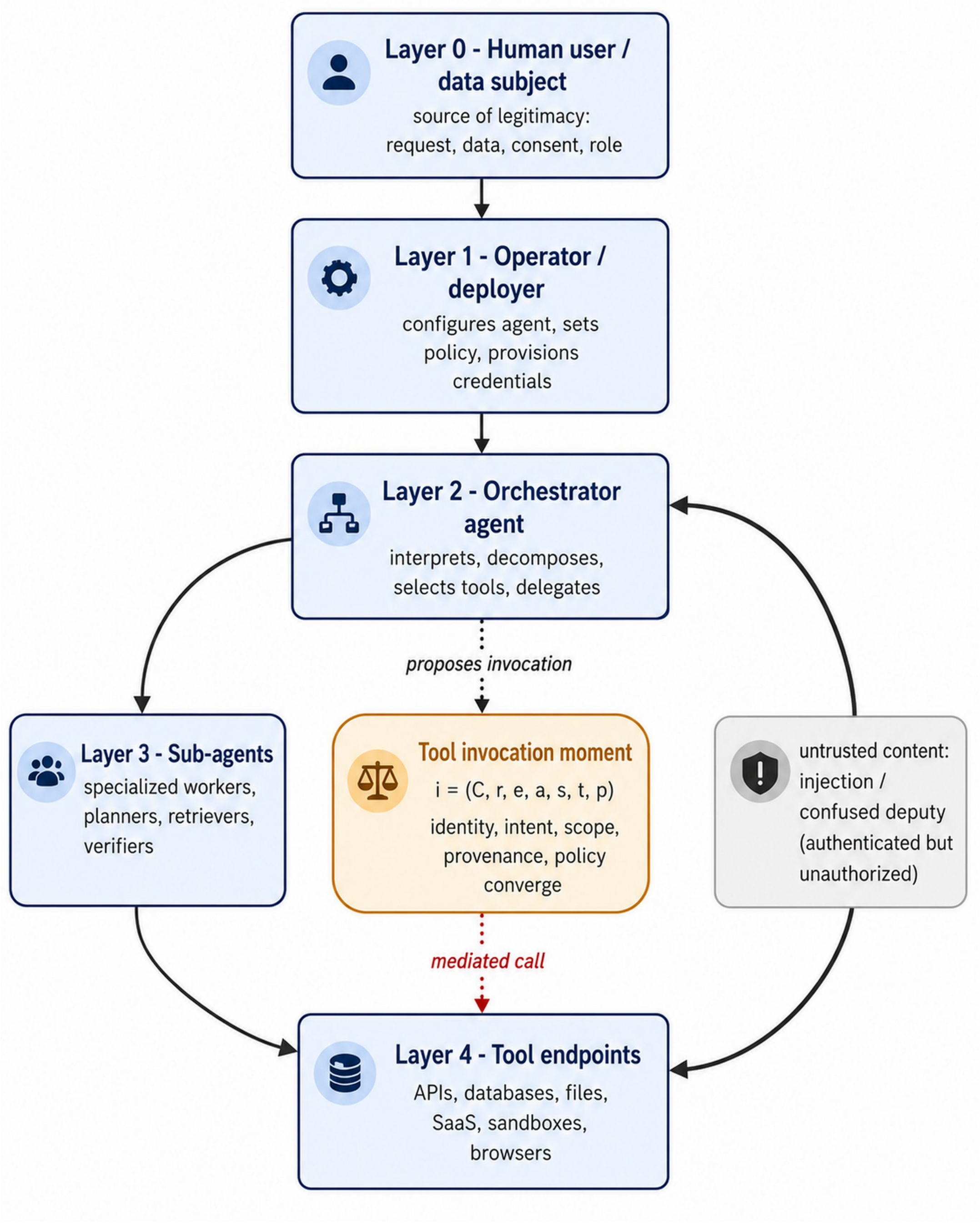

**Fig. 2** The principal hierarchy and the tool invocation moment

### 2.4 Authentication, authorization, and access control

These terms are often conflated; however, we distinguish between them because each maps to a different failure mode. Authentication establishes the identity of a human, service account, runtime, subagent, or tool server. It is necessary but insufficient: knowing that a request came from a particular agent does not establish that the agent can perform the action on behalf of a human or task. Authorization determines whether a requested action is allowed under policy, scope, delegation, consent, or entitlement; an agent may hold a valid token but may lack sufficient delegated authority for a specific request. Access control is a model and mechanism used to express and enforce authorization. Role-based access control (RBAC) assigns permissions to roles [22]; attribute-based access control (ABAC) evaluates the attributes of subjects, objects, actions, and environments [23]; relationship-based access control (ReBAC) bases authorization on relationships such as ownership or membership [24]; and policy-based access control (PBAC, used here to mean explicit policy evaluation over contextual facts and obligations rather than the "purpose-based" interpretation found in some literature) completes the model space. Capability-based security treats unforgeable references as authority, with attenuation restricting the delegated authority [10,25]. The implication is that an LLM's internal decision to call a tool is not itself an authorization decision: the model proposes; the system must still authenticate principals, decide authorization, enforce at an appropriate point (a *policy enforcement point*, PEP, distinct from the policy decision and administration points, PDP and PAP), and record provenance. Two further terms recur throughout: a *non-human identity* (NHI) is a digital identity for a non-person entity, such as a workload, service account, or agent; and *attenuation* is the narrowing of authority (by scope, time, resource, or action) as it is delegated [10,25].

### 2.5 Review methodology

This is a structured narrative review, appropriate where a field is conceptually fragmented, terminology is unstable, and the goal is synthesis and research-agenda formation rather than meta-analysis [26,27]. It does not present a PRISMA flow diagram and should not be read as claiming systematic review coverage; instead, it uses an explicit search strategy, inclusion criteria, and coding scheme. The review was guided by five questions corresponding to the paper's layers: how agents should be identified as non-human principals, how authority should propagate across the hierarchy, where runtime enforcement can be placed, how prompt injection functions as an authorization failure, and what provenance is required to attribute terminal actions. Sources were drawn from ACM, IEEE, Scopus-indexed literature, Google Scholar, arXiv (with peer-reviewed version checks), and publisher proceedings, using terms combining agent tool use, agentic authorization, agent/non-human identity, delegation, prompt injection, runtime policy enforcement, and agent audit/provenance, with backward/forward citation searches for foundational concepts. The primary window for the LLM-agent literature was 2023–2026; earlier work was included, supplying foundational concepts. Each work was coded by year, system layer, mechanism family, threat class, and evidence type, where evidence type distinguishes peer-reviewed work with experimental evaluation, peer-reviewed conceptual work, research preprints with and without evaluation, published standards, draft specifications, and industry or association reports. That classification is carried through the comparative tables and the gap analysis, and is reported explicitly in Appendix A, so that a claim resting on a conceptual preprint is not read as equivalent to one resting on an evaluated, peer-reviewed system. The initial search yielded approximately 180 candidate works; after applying the inclusion criteria (authorization-relevance, peer-reviewed or recognized preprint venue, English language, and non-vendor-marketing), 89 sources were retained as primary references. The remaining screened works provided background understanding but either duplicated coverage of a mechanism already represented by a stronger source, fell outside the authorization-specific scope, or were vendor reports that could not be independently corroborated. The full search and coding process, including the databases, search concepts, screening criteria, and the rule by which a gap-analysis cell is rated as having no coverage, is given in the Supplementary Material.

### 2.6 Relationship to existing surveys

Existing surveys cover LLM agent capabilities, architectures, planning, tool use, and evaluation [18–20], and security-oriented studies have demonstrated indirect prompt injection in LLM-integrated applications [15]. These are complementary but do not treat authorization at the tool invocation moment as a central concern.

**Table 1. Positioning relative to adjacent research.**

| Work | Emphasis | Gap addressed here |
|---|---|---|
| [18] | Tool/retrieval augmentation as an architectural pattern. | It does not focus on identity, delegation, or runtime authorization. |
| [19] | Construction and evaluation of autonomous agents. | Security is not organized around the principal hierarchy or the time of invocation enforcement. |
| [20] | Tool use, retrieval-augmented generation (RAG), planning, and feedback learning. | It does not synthesize access control models, delegations or auditability. |
| [15] | Indirect prompt injection attacks. | Attack-focused; no authorization architecture across identity, delegation, enforcement, or auditing. |

The present survey's contribution is integrative: access-control research explains how decisions can be expressed and enforced; capability and confused deputy research explains why authority should be attenuated; prompt injection research explains how untrusted content corrupts the authority chain; and provenance research explains how actions can be reconstructed. The central claim is that tool invocation is safe only when all these conditions align.

### 2.7 Threat and adversary model

The threat model is expressed against the principal hierarchy, which crosses authority boundaries, not merely network boundaries, between the user and operator, deployment and planning, orchestrator and subagent, and runtime and tool endpoints. The adversary may occupy several positions: a careless or malicious user issuing instructions beyond their authority (Layer 0); an external attacker controlling content returned by a tool (an email, web page, document, ticket, or retrieved memory) and creating an indirect injection path (Layer 4) [15]; a compromised sub-agent inducing the orchestrator to act outside the delegated task (Layer 3); a malicious tool provider misrepresenting behavior through names, descriptions, or schemas; a network attacker intercepting or replaying calls; or an insider provisioning overly broad credentials.

Six threats are in scope: (1) unauthorized invocation: calling a tool without sufficient authority; (2) privilege escalation: authority expanding rather than attenuating down the hierarchy; (3) confused-deputy failures: legitimate authority induced to act for an unauthorized party [10]; (4) prompt injection as authorization bypass: untrusted content causing the agent to override the instruction hierarchy [15]; (5) data exfiltration: combining permitted read access with a transmission tool; and (6) audit failure: a terminal action that cannot be reconstructed back to a human sponsor, policy, scope, and decision. Out of scope are model-weight tampering, training-time manipulation, and hardware compromise, except where they directly affect runtime authorization; persistent-memory and poisoned-retrieval attacks are in scope, where they influence subsequent invocation. The working assumption is not that LLMs are malicious but that they are unreliable authorization subjects; the model output should be treated as an input to authorization, not as the authorization decision.

## 3. Agent Identity and the Non-Human Identity Lifecycle

If an agent cannot be named, authenticated, scoped, monitored, and revoked as a distinct non-human identity (NHI), later mechanisms cannot determine whose authority is being exercised over the data. Therefore, identity is a part of the security semantics of every tool call. The central claim of this section is that agent identity is a lifecycle problem, not a credential format issue. Static API keys, OAuth tokens, OpenID Connect (OIDC) assertions, proof-of-possession mechanisms, workload identities, decentralized identifiers (DIDs), and verifiable credentials (VCs) each provide building blocks, but none alone answer how authority should bind to a changing, partially autonomous runtime acting for a human sponsor, under operator policy, through subagents, and across organizations.

### 3.1 What distinguishes AI-agent identity

Enterprises have long used NHIs, including service accounts, workload identities, bots, robotic process automation (RPA) accounts, and application programming interface (API) clients. The novelty of this approach lies in the fact that tool-using agents combine NHI properties with runtime interpretation, planning, and dynamic capability selection. A conventional service account runs a stable program against predefined resources; an agent translates goals into calls, chooses among tools at runtime, incorporates untrusted outputs into later decisions, and delegates subtasks [2,3,19]. Four properties distinguish agent identity: interpretive autonomy (the agent decides which operation appears relevant; therefore, credentials cannot be bound to an application name alone), ephemerality and compositionality (short-lived planners, workers, and tool agents whose actions have durable effects), dynamic capability acquisition (the effective action space depends on the tool registry, not only model weights), and multi-hop traversal (a request passes through the orchestrator, sub-agents, and several endpoints before an external action). Therefore, a useful identity record identifies the agent instance, operator, authorizing human or process, available toolset, delegated scope, temporal validity, and evidence needed to reconstruct the later actions.

### 3.2 Credential types and their security properties

Credential mechanisms differ in terms of what they prove, how narrowly they are scoped, and how easily they can be revoked for security reasons. Static API keys are bearer secrets: possession suffices, and the endpoint cannot distinguish the intended agent from a compromised runtime or injected call. Empirical studies have shown that such secrets are frequently mishandled [28,29]. OAuth-family delegated tokens separate resource owners, clients, authorization servers, and resource servers, which are mapped to agents acting on behalf of users. However, ordinary scopes such as `email.send` do not encode the natural language task, recipient, content boundary, sponsor or delegation chain. Formal analysis shows that OAuth guarantees hold only under a precise attacker model and careful role separation [30]. OpenID Connect assertions authenticate the end user [31] but must not be mistaken for authorization to perform a tool action; an agent may authenticate the user while exceeding the user's delegated purpose. Proof-of-possession and channel-bound tokens reduce replay by requiring key control but do not prove that an LLM selects a tool for an authorized purpose. Workload identity authenticates the runtime, which is valuable in cloud-native deployments. However, a resource server still needs to know whether the runtime acts for a specific user, task, scope or window. DIDs and VCs offer portability for cross-organizational identity, expressing claims about an agent, owner, or delegation; however, their practical use for tool invocation is immature, with open problems in issuer trust, revocation, selective disclosure, and IAM integration [32,33].

**Table 2. Credential mechanisms for tool-using agents.**

| Mechanism | Scope granularity | Revocability | Multi-hop support | Key limitations for agentic use |
|---|---|---|---|---|
| Static API key | Coarse (endpoint/account) | Manual, irregular | Weak | Cannot bind task purpose, sponsor, temporal validity, or sub-agent chain |
| OAuth delegated token | Scope/audience/resource | Deployment-dependent | Partial (token exchange) | Scopes too coarse for natural-language tasks [30] |
| OIDC identity assertion | Identity claims | Provider session | Identifies human, not full chain | Authenticates user but does not authorize the terminal action [31] |
| Proof-of-possession token | As delegated, + possession | Token/key lifecycle | Protects hops from replay | Proves possession, not semantic task authorization |
| Workload identity | Workload/environment | Short-lived, rotatable | Strong service-to-service | Identifies runtime; needs user-delegation context |
| DID/VC credential | Fine-grained, claim-based | Ecosystem-dependent | Promising cross-org | Trust governance, revocation, policy interpretation immature [32] |

No credential type is sufficient in isolation. A strong agent identity requires composition: workload identity to authenticate the runtime, delegated authorization to bind resources, possession proofs to reduce replay, structured claims to represent task constraints, and provenance to preserve the chain across the hops. The unresolved problem is making this composition enforceable at the latency and dynamism of the agentic workflow levels.

### 3.3 Dynamic and just-in-time provisioning

Pre-provisioning broad credentials creates standing privileges, which are poorly matched to agents that do not know in advance which tool, object, or sub-agent a task requires. Just-in-time (JIT) provisioning issues or elevated authority only when a specific action requires it under contextual constraints, consistent with the zero-trust principles of continuous verification and least privilege [34]. Provisioning should be driven by the proposed invocation : the system asks whether the chain  is valid, the operation  on  is consistent with the scope , the arguments  satisfy the policy, the temporal context permits the action, and provenance  is sufficient to perform the action. This transforms the issuance process into an authorization-dependent event. The main design challenge is the bootstrap paradox, that is, how an agent obtains a credential before it has one. This is resolved by separating the *bootstrap identity* (authenticating the workload) from the delegated action authority (issued only after evaluating the user, policy, and requested call) so that the bootstrap credential does not become a universal secret. The remaining gap is semantic; existing mechanisms can carry structured facts, but the ecosystem lacks a widely adopted vocabulary for representing human-approved task purposes, agent roles, tool boundaries, and delegation chains in a form that resource servers can enforce.

### 3.4 Lifecycle: provisioning, registration, rotation, revocation

A lifecycle view begins before the credentials are issued. Provisioning establishes an owner, deployment environment, business purpose, permitted tool inventory, risk class, logging requirements, and default-privilege boundaries. Registration makes identity governable: a unique identifier, owner, operator, creation time, issuer, allowed environments and users, allowed tools, and policies, plus agent-specific fields such as model/runtime, system-prompt authority, memory stores, retrieval sources, sub-agent creation rights, and human approval thresholds. Rotation must account for both the credential lifetime and agent behavior; the strongest pattern combines a stable identity record with a short-lived, narrowly scoped authority, although other lifecycle designs may suit different deployment contexts. Revocation is more difficult than human offboarding; triggers include task completion, workflow cancellation, tool removal, prompt or model change, memory poisoning, credential leakage, abnormal call patterns, and detected injections. A four-state model (*registered, active, suspended, and retired*) prevents conflating revocation with deletion and preserves audit records for nonrepudiation. Lifecycle governance must also treat tool-set drift: if an agent approved to read a calendar later gains access to an email, file, or payment, or if a tool server changes its descriptions, the agent's accessible action space expands without a code deployment; therefore, tool addition, schema change, and sub-agent delegation should require reapproval or rescoping.

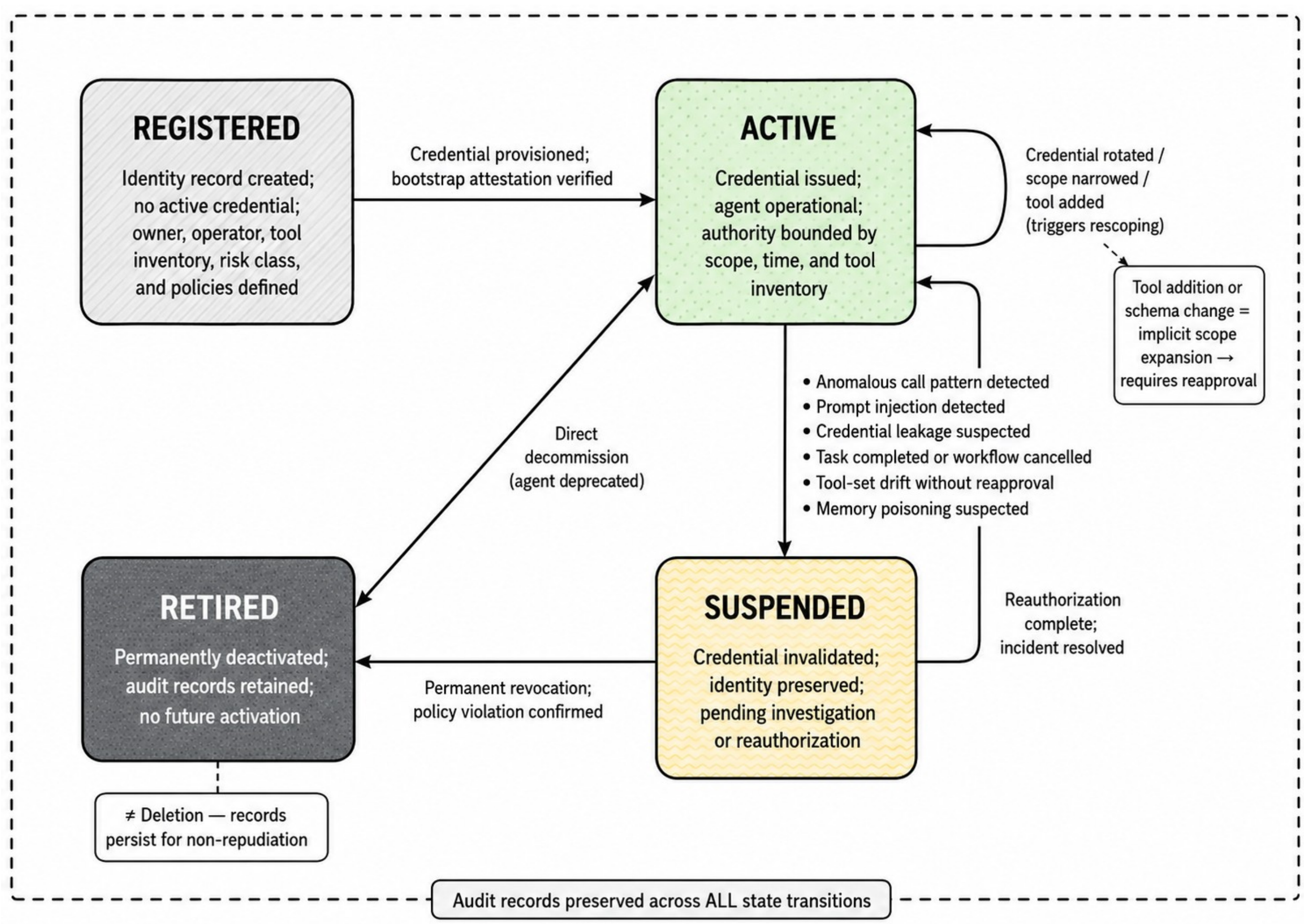

**Fig. 3** Non-human identity (NHI) credential lifecycle state diagram showing the four states and transition triggers

### 3.5 Cross-boundary identity and the shadow-agent problem

This problem is exacerbated across organizations: a workflow may span a user in one organization, a platform operated by another, a third-party tool server, and a resource server that enforces its own policies. A resource server may trust the platform but not know the human sponsor or receive a delegated token that proves tool access but not the complete delegation chain. DIDs and VCs can express portable claims, but self-sovereign identity (SSI) literature shows that interoperability, governance, revocation, and trust framework design remain open [32,33]. Unresolved questions include issuer trust (who may assert an agent's role or capability), revocation propagation (how quickly partners learn that prior claims are invalid), scope translation (mapping one organization's task policy to another's API scopes), privacy minimization (how much of the chain to expose to an endpoint), and liability (whose evidence assigns responsibility when a multiagent chain causes harm).

A related risk is the shadow agent, which is an agentic workflow, assistant, script, or automation that can invoke tools but is not registered, owned, or governed by the organization itself. Shadow agents create unmanaged paths through the principal hierarchy using human-delegated grants, shared keys or credentials embedded in a local environment. Because they are absent from the inventory, downstream systems may only see a generic key, making attribution impossible. The remedy is to extend the inventory from credentials to capability graphs that record which tools an agent can call, which data sources can influence its context, which sub-agents it can create, and which principal is accountable. Heavyweight governance drives agent creation underground, and the practical challenge is low-friction registration and least-privilege credential vending, which makes the governed path easier than the unmanaged path.

The identity-layer open problems (an agent identity schema, bootstrap attestation outside controlled infrastructure, scope representation for natural-language tasks, multi-hop revocation, and cross-boundary trust frameworks) are taken up with the cross-cutting challenges in Section 9. Identity is necessary but insufficient: a well-identified agent can still misuse valid authority if the delegation is too broad, which is the subject of Section 4.

## 4. Delegation, Scope Propagation, and the Confused Deputy

A well-identified agent can still misuse valid authority if that authority is too broad, the context is lost across the steps, or a downstream component exercises authority for an unapproved purpose. This section shifts from *the agent to the manner in which authority travels*. In the invocation tuple, delegation concerns the chain , scope , temporal context , and provenance . A delegation mechanism is adequate only if it preserves the authority chain while narrowing the scope of what each downstream actor can do. The common failure is authority drift: the effective permission at a tool endpoint becomes broader, less attributable, or less time-bound than the authority granted at the start of the task, which is the thread connecting transitive delegation failures, overprivileged tools, confused-deputy attacks, aggregation inference, and temporal decay.

### 4.1 Delegation models: implicit vs. explicit, scoped vs. unscoped

Classical delegation models distinguish who can delegate, what can be delegated, whether authority is granted or transferred, how depth is bounded, and how delegation is revoked [35,36]. These are directly relevant because a tool-using agent is a delegate that receives tasks and converts them into concrete task calls. However, the difficulty lies in the fact that delegation in agent workflows is often implicit. A user who writes "triage these tickets" or "summarize and send the report" expresses a goal, not a machine-checkable *authorization envelope* (a structured artifact that explicitly bounds the permitted operations, objects, recipients, time window, and delegation depth for a task); the model infers a chain of operations, creating a semantic gap between the intent and what the system can enforce. OAuth 2.0 provides the deployed baseline, representing permissions through scopes, lifetime, audience, and authorization-server policy [37], but scopes such as `files.read` do not encode task purpose, permitted recipients, the relevant document set, invocation limits, or task-completion expiry.

Recent extensions help but do not close the gap: Rich Authorization Requests (RAR) add structured, transaction-specific authorization details [38], and token exchange supports on behalf of patterns across services [39]. The unresolved problem is not the token format but how to bind a natural language task, a chain of agentic decisions, and a terminal action into an enforceable authorization object.

Two axes organize the design spaces. Implicit vs. explicit: Implicit delegation infers authority from instruction or session context, whereas explicit delegation produces a concrete artifact (token, capability, credential, policy decision, or task-scoped envelope). Scoped versus unscoped: Scoped delegation constrains resources, operations, arguments, time, purpose, and downstream delegation; unscoped delegation grants ambient authority that is reusable across tasks. Agent-safe delegation should move toward the explicit-and-scoped quadrant of this design space. The emerging agent-identity literature agrees: South et al. [40] argue that agent delegation should be authenticated, authorized, and auditable, and that natural language permissions must be translated into reviewable access-control configurations. The AI Agent Authentication and Authorization (AIMS) draft treats agents as workloads with delegated authority while emphasizing that user and system contexts must be preserved for authorization and audit [41].

### 4.2 Multi-hop chains and the transitive delegation problem

Multi-hop delegation is structurally more complex than single-hop delegations. At each hop, the context may be compressed, transformed, or lost: the orchestrator summarizes the user's task, gives a sub-agent only a partial description, and the sub-agent calls a tool with a credential identifying the runtime but not the user. The resulting action is authenticated but semantically detached from the original authorization: the transitive delegation problem. Transitivity is not inherently unsafe, but in agentic systems, transitive authority often lacks explicit attenuation: a retrieval agent given a "summarize" subtask should not inherit write privileges; a browser agent authorized to read one page should not gain authority to submit forms or exfiltrate data. Tallam [42] frames authorization propagation as a workflow-level problem that cannot be reduced to endpoint RBAC/ABAC/ReBAC checks. Tomašev et al. [43] argue that AI delegation involves authority, responsibility, accountability, and trust relationships, and that delegation is an *authority transformation*, not mere routing.

Capability-style mechanisms support this attenuation effect. Macaroons carry contextual caveats restricting how, where, when, and by whom authority may be exercised [44]; therefore, an orchestrator can derive a narrower capability for a subagent by adding a task identifier, endpoint, use limit, time bound, or recipient constraint, provided that the caveats are meaningful to the verifier. Invocation-Bound Capability Tokens (IBCTs) make this agent-specific: the Agent Identity Protocol (AIP) combines identity, attenuated authorization, and provenance binding into a token chain for MCP and Agent-to-Agent (A2A) interactions; thus, each hop carries a bounded capability and evidence of its derivations [45]. These are early proposals; their feasibility depends on standardized claim vocabularies, efficient verification, privacy-preserving disclosures, and adoption by the tool servers. These proposals converge on a common position, though it should be read as a shared design intuition among a small number of early sources rather than as a settled consensus: multi-hop delegation should preserve the chain, attenuate scope at each hop, bind authority to a task or invocation, and produce audit evidence. What remains unsettled is how to represent the task scope precisely enough for machines to understand and for users to comprehend it.

### 4.3 Least privilege in agentic systems

Least privilege [8] is both more important and harder to achieve for agents because their value lies in flexible execution, which tempts developers to expose broad tool inventories and credentials at initialization. The least privilege must be defined at several levels: tool-inventory (which tools are visible), invocation (whether the current step requires the selected tool), argument (the object, recipient, query, path, amount, or payload), data (how much output is returned to the model, with sensitive fields redacted), and delegation (preventing a sub-agent from receiving broader authority than its subtask requires). Recent systems have shifted from static provisioning to runtime narrowing. For example, AgenTRIM reconstructs and verifies tool interfaces offline and then enforces per-step least privilege through adaptive tool filtering and status-aware validation [46], which is evaluated on AgentDojo adversarial tasks [47]. PAuth targets the task scope side, deriving precise task-scoped envelopes from natural language slices and directly addressing the intent-permission gap [48]. However, task slicing can be ambiguous, and a manipulated prompt may request a slice that is broader than intended. The field lacks agreed evaluation criteria; a least-privilege mechanism should be measured not only by the attack prevention rate but also by false denials, user burden, latency, IAM interoperability, and audit preservation.

### 4.4 The confused deputy in multi-agent systems

The confused deputy problem [10] is the clearest lens for agentic authorization failures: a privileged program is tricked into using its authority for another party because it cannot distinguish which of its multiple authorities should be applied to the situation. Agents recreate this at scale by holding operator credentials, acting as users, reading untrusted content, and blending these into one context. The agentic case differs from the OS case in three ways: the decision procedure is prompt-conditioned and probabilistic; the authority chain is distributed across humans, operators, runtimes, sub-agents, and services; and the adversarial request arrives as content embedded in a document, email, or tool response rather than as an explicit API call.

A worked path (Figure 4): (1) a user authorizes an orchestrator to summarize invoices from an inbox; (2) the orchestrator reads a message containing untrusted text; (3) the text requests a side effect, such as forwarding data or initiating a payment; (4) the orchestrator passes it to a finance sub-agent without preserving the read-only scope; (5) the downstream tool receives a valid credential and executes the write; (6) the terminal service authenticates the caller but cannot determine that Layer 4 content caused Layer 2 to exceed the authority delegated by Layer 0 and constrained by Layer 1.

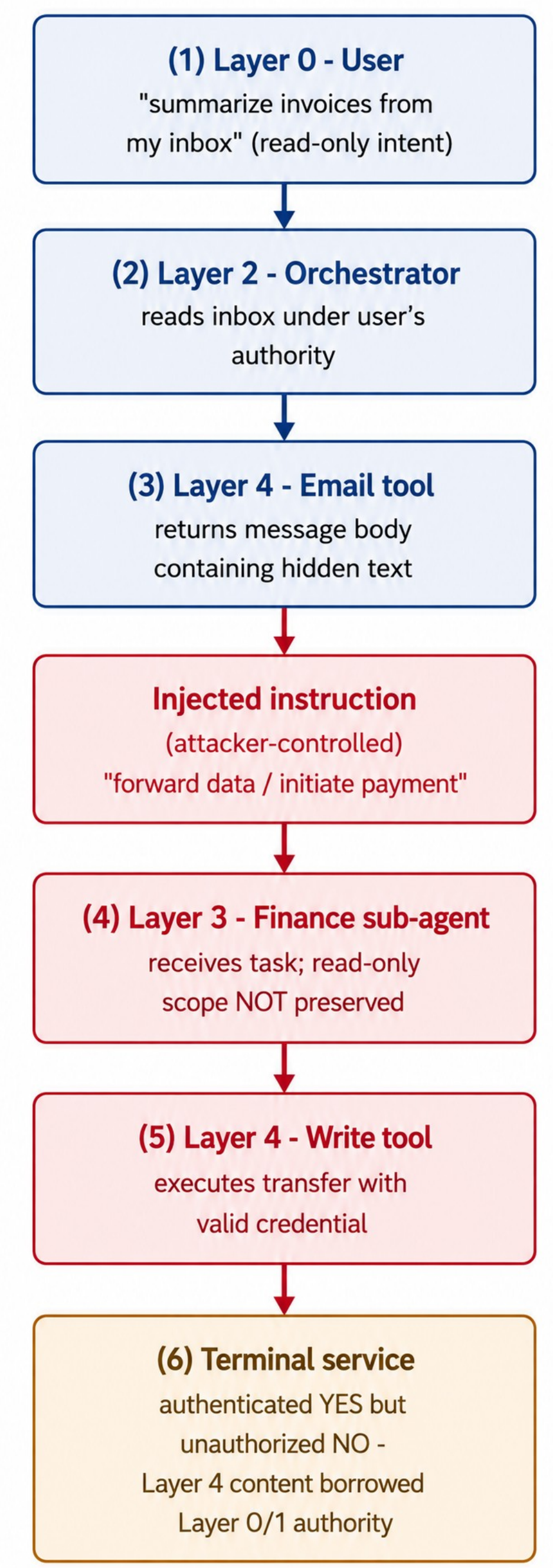

**Fig. 4** A confused deputy path through the principal hierarchy

### 4.5 Aggregation inference across individually authorized tools

Per-tool least-privilege does not prevent unauthorized *aggregate* knowledge. The classical database inference problem, in which tracker attacks leak protected information even when individual queries appear safe [49,50], provides a structural analogy, although the threat model differs: database inference assumes a fixed dataset with statistical queries, whereas agent aggregation involves dynamic, heterogeneous tool outputs composed through natural-language reasoning. The analogy holds at the *principle* level (individually authorized accesses can produce unauthorized knowledge when combined), but the *mechanism* level differs (query auditing and noise injection do not translate directly to multi-tool agent workflows). Agents exacerbate this problem because they are designed to integrate context across sources. For example, a personal assistant combining calendar, location, email, and contacts may infer relationships that no single tool disclosed. An enterprise agent combining records, tickets, messages, and commits may infer commercially sensitive information from these data points. Tallam [42] identifies aggregation inference as a core subproblem of authorization propagation, implying that scope  cannot be interpreted only as independent tool permissions; it must also constrain cumulative information use, derived outputs, and cross-tool composition. This pushes authorization toward provenance- and information-flow-aware designs and exposes a usefulness-privacy tension: a more capable synthesis derives more facts that are not present in any single information source. Current literature offers directions rather than settled answers (provenance tracking, data-flow labels, query budgets, purpose limitation, declassification rules, and human approval for high-risk derived outputs); however, the lesson is clear: least privilege must apply to workflows and derived information, not only to isolated calls.

### 4.6 Temporal validity and authorization decay

Delegated authority is also scoped by time, which is important because agent workflows can be long, asynchronous, or partially autonomous. Authorization decay occurs when an authority valid at one point becomes stale, excessive, or ambiguous as the workflow evolves, the task is completed, the user changes instructions, the relevant object changes sensitivity, or the context is corrupted. Token expiration mitigates only part of this: a token may be unexpired while the task is complete or expires mid-workflow and forces reauthorization without reconstructing the purpose. Temporal context  should therefore include token lifetime, task lifetime, approval time, last policy check, workflow state, execution count, and revocation status [41,42]: a call valid as a task's first step may be invalid as its fiftieth step after multiple delegations. Defenses, invocation-bound credentials, maximum-use constraints, time-limited task envelopes, reauthorization checkpoints, event-triggered revocation, and human confirmation for high-consequence actions must be calibrated, as excessive reauthorization breeds fatigue and broad "remember my approval" grants, whereas insufficient reauthorization permits stale authority.

### 4.7 Delegation flow archetypes

Three architectural archetypes organize the design space and may be combined with risk [41,51]. In agent-mediated delegation, the agent carries delegated authority to the tool server, which is operationally simple and supportive of autonomy, but is most exposed to confused-deputy and ambient-authority failures; it is appropriate for low-risk tasks but dangerous with broad write authority and untrusted outputs. In owner-mediated delegation, human or policy authority remains involved through just-in-time approval, consent screens, or step-up authorization, which is attractive for high-consequence actions but limited by usability, depending on concise, reviewable representations of the requested action. In server-mediated delegation, the resource server or gateway evaluates both the agent and user context at the point of access, which is closest to complete mediation, able to enforce object-level policy, and reject calls lacking principal-chain evidence, but requires standardization across agents, identity providers, gateways and servers. A useful pattern is risk-tiered composition: low-risk reads use agent-mediated with narrow tokens and logging; medium-risk operations require server-mediated checks; and high-risk actions (external messages, permission changes, payments, deletions, and deployments) require owner-mediated reapproval.

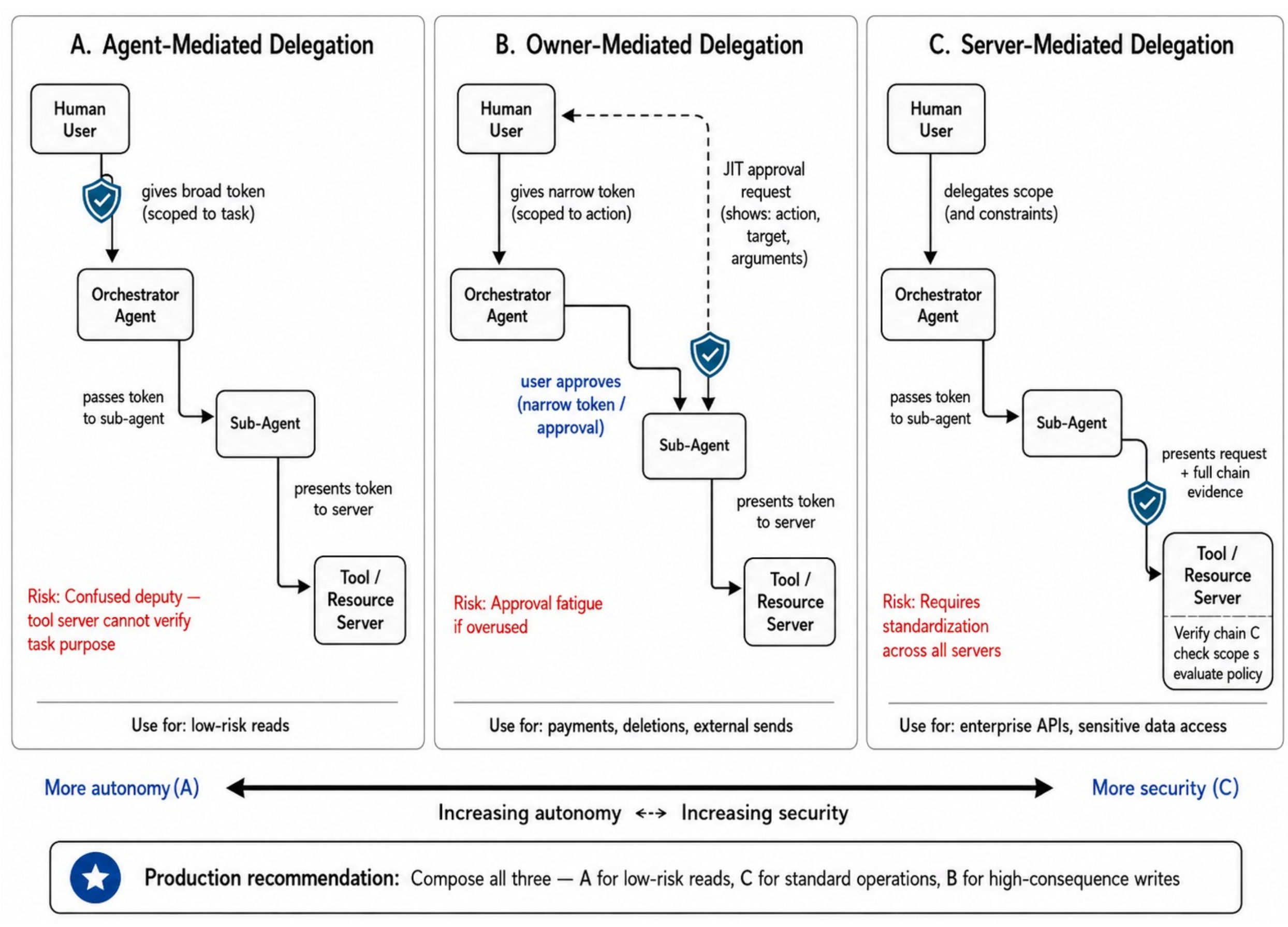

**Fig. 5** Delegation flow archetypes: agent-mediated, owner-mediated, and server-mediated, showing the trade-off between autonomy and security

One pattern runs through this section: every delegation failure mode is countered by making authority explicit and narrowing it at the point of use, whether the failure is scope widening, context loss, confused deputy, aggregation inference, temporal decay, or cross-boundary ambiguity. Delegation is therefore a continuing property of a workflow rather than a single token issuance: the agent must obtain, preserve, narrow, and evidence authority as execution unfolds. Whether those semantics are actually verified at the invocation layer is the subject of Section 5.

## 5. Runtime Enforcement at the Tool Invocation Layer

A principal chain can begin with a legitimate request, pass through a correctly identified orchestrator, and still end in an unauthorized action if the delegated authority is not checked at the point of use. Runtime enforcement evaluates an attempted invocation *before* an external effect is triggered. The central claim is that a prompt, system message, or alignment policy is not a policy enforcement point (PEP): the model may propose an action, but enforcement requires independent control that can mediate, deny, or transform the call, request authorization, and produce audit evidence. This is complete mediation [8], which is more difficult because the action is selected by a probabilistic planner rather than by a fixed route. This position should not be read as dismissing model-level progress: instruction hierarchy training [52] and preference optimization for security alignment [53] demonstrably reduce the rate at which models propose unauthorized actions, serving as a valuable *complementary* layer that reduces the load on external enforcement efforts. However, they cannot serve as the *sole* enforcement mechanism because they provide probabilistic compliance rather than deterministic mediation, cannot be formally verified against arbitrary adversarial inputs, and do not produce audit evidence of a policy decision. A further reason is specific to how these models are trained. Agentic systems optimize objectives that are proxies for what their designers intended, and a survey of reward hacking in agentic LLM systems documents how such systems come to satisfy the measured objective while defeating the intended one [54]. An agent's internal objective is therefore not a dependable proxy for its principal's authorization, which is an argument for placing the authorization boundary outside the component being optimized rather than inside it. The appropriate architectural model treats model-level authorization as the first filter (reducing the frequency of enforcement challenges) and external PEPs as the authoritative decision layer.

### 5.1 The runtime enforcement gap

The gap is the separation between the policies stated at the design time and the controls applied when an agent invokes a tool. Many systems can state broad intentions (use only approved tools, do not exfiltrate data, ask before sending), but lack a mechanism that deterministically intercepts the terminal operation and evaluates it against the task-specific authority chain. Authentication does not close this: a valid token proves that the request comes from a known runtime, not that the operation matches the current task, scope, recipient, data class, or window; and tool selection by the model is insufficient because the model context may contain instructions from the user, operator, documents, web pages, and prior outputs. Analyses of tool-enabled agents have identified overprivileged tools, ambient authority leakage, and capability-intent mismatches as recurrent causes [11,12].

Three recurring patterns have emerged. Front-loaded authorization approves a broad tool or account before work begins, which is weaker than coarse OAuth consent because the call sequence is unknown at the time of consent. Inventory-only restriction reduces the number of visible tools but leaves each broadly empowered, ignoring whether a permitted call's arguments are safe. Post-hoc monitoring logs after execution support audits but cannot prevent irreversible side effects. Existing cloud-provider IAM systems represent the closest deployed approximation to invocation-layer enforcement: AWS IAM policies on Bedrock Agents provide action-group-level permissions that restrict the tool categories that an agent can invoke; Google Cloud Vertex AI Agent Builder supports OAuth-scoped tool access with per-tool credential binding; and Azure Managed Identity with RBAC provides workload-authenticated, role-scoped access to downstream services. These systems satisfy portions of R1 (workload identity authentication) and provide coarse R6 enforcement (tool-category restrictions), but they do not address task-purpose binding (the agent's credential does not encode *which user task* justifies the call), delegation-chain preservation (the downstream service sees the agent's identity but not the human sponsor's delegated scope), natural-language scope constraints (permissions are expressed as API-level actions, not task-level intents), or path-aware enforcement (whether the current call is safe given the sequence of prior reads and writes). Therefore, they close the *infrastructure* portion of the gap while leaving the *semantic* portion (the distance between "this agent can call this API" and "this specific invocation is authorized for this task on behalf of this user") open.

Recent systems have shifted to runtime mediation: Progent represents privileges as symbolic policies over tool names and arguments and checks every call during execution [55]; AgenTRIM applies adaptive runtime filtering grounded in a verified interface [46]; and AgentBound enforces declarative policies at the MCP execution boundaries without server modification [12]. They differ in placement and expressiveness but share the assumption that control must be applied to specific actions. These systems agree that enforcement should be both invocation-aware (checking the specific call) and path-aware (considering the prior sequence of reads and delegations); the latter is needed for cases such as "read a sensitive file, then send an email," where neither operation is forbidden in isolation. Information flow control (IFC) approaches make this explicit by tracking how private or untrusted information moves across tools, models, memory and outputs [13,56]. Therefore, the gap is a missing control plane that connects action mediation, data flow tracking, delegation states and policy evaluation.

### 5.2 Access-control models for dynamic agent environments

No classical model maps cleanly onto these agents, and each fails in an instructive way. RBAC is administratively simple [22] but role membership is too coarse for per-invocation decisions: a "support assistant" role spans permissions that should vary by user, customer relationship, ticket state, and sensitivity. ABAC is more expressive, evaluating subject, object, action, and environment attributes [23], and all of agent identity, sponsor, purpose, object classification, recipient, time, and provenance source are relevant attributes; its quality, however, depends on trusted attribute capture, and where the runtime cannot reliably supply purpose, sensitivity, or untrusted-source status, an ABAC decision either over-denies or rests on attributes produced by the very model it governs. ReBAC authorizes on relationships such as ownership or assignment [24], which suits agents acting inside collaboration systems but cannot express whether a multi-step sequence stays within an approved purpose. Capability-based security couples designation with authority and supports attenuation [25], which maps well onto sub-agents receiving narrowed references rather than ambient credentials, though a capability at the granularity of "email API" still over-authorizes; capabilities are strongest when invocation-bound, object-specific, time-limited, and auditable. Mandatory access control and information-flow control address the remaining question of whether information from one source may influence an action at another sink: Denning's lattice model grounds secure flow across security classes [57], SEAgent applies it to privilege escalation through information-flow graphs [58], and GAAP tracks how private user data are accessed and disclosed even when the model provider is untrusted [13].

The object-capability (ocap) discipline deserves separate mention because it differs in kind rather than degree. Where the PEP model detects and blocks misuse of ambient authority, ocap eliminates ambient authority structurally: a component can exercise only the authority granted through a reference it holds, so there is no channel through which untrusted content can borrow unrelated authority and the confused deputy cannot arise. Some agent-security designs follow ocap principles implicitly when they pass narrowed references between sub-agents instead of sharing credentials, but no framework in the reviewed corpus enforces the discipline end-to-end across the principal hierarchy, and its integration with natural-language task descriptions is unexplored.

The practical conclusion is that agent authorization is hybrid: RBAC for broad entitlements, ABAC for invocation context, ReBAC for relationships, capabilities for attenuated delegation, MAC/IFC to keep untrusted or private data from privileged sinks, and ocap principles to structurally eliminate confused-deputy paths where feasible, all composed through a policy engine. The challenge is to compose them without opaque policies or intolerable latency.

### 5.3 Policy enforcement points in the agent stack

The question for agents is less "what is the policy language?" than "where can the PEP see the relevant context and still block the action before execution?" There are at least six placements, each seeing different facts and failing differently. The proposed model-adjacent guard constrains tool calls: low cost, sees names and arguments, but may not see the resource state, and can be bypassed if tools are called off the mediated path. A runtime orchestrator guard mediates planning, selection, and subagent calls, which provides more workflow context but is framework coupled. A tool adapter wrapper enforces argument-level constraints independently of the model, which is practically deployable but is blind to hidden side effects. A tool server/gateway PEP enforces resource and token policies inside or around an MCP server or API gateway, which is mature for APIs but may not be familiar with natural language tasks. A sandbox/OS boundary restricts file, network, and process access, which is mature but controls host effects and not application-level purposes. A service-side resource PEP enforces ordinary API authorization: mature but often sees only the credential holder and not the entire chain.

**Table 3. Runtime PEP architectures.**

| Placement | Granularity | Latency | Multi-hop awareness | Main limitation |
|---|---|---|---|---|
| Model-adjacent guard | Tool name, schema, arguments | Low | Weak unless trace supplied | Bypassable if tools called off the mediated path |
| Orchestrator/ runtime guard | Step, task, sub-agent, trace | Low–moderate | Moderate–strong within one framework | Framework-specific |
| Tool-adapter wrapper | Tool and argument | Low | Limited unless claims forwarded | Blind to hidden side effects |
| Tool-server/ gateway PEP | Endpoint, object, method, token claims | Moderate | Moderate if tokens carry chain | May not know task or untrusted-source influence |
| Sandbox/OS boundary | Resource class, process capability | Moderate | Weak semantically | Controls host effects, not app-level authorization |
| Service-side resource PEP | Resource and operation | Low–moderate | Weak unless claims preserved | Often sees only the credential holder |

The recurring tradeoff: the closer the PEP to the model, the richer the task context, but the weaker the enforcement authority; the closer to the resource, the stronger the boundary, but the weaker the semantic context. AgentBound is enforced at the MCP boundary [12], Progent applies symbolic tool-call policies without modifying the internals [55], and AgenTRIM grounds runtime filtering in a reconstructed interface [46]. Formal rule-checking constraints permit actions through explicit security rules rather than probabilistic detection [59]. Sandboxing is necessary but incomplete; it may stop `rm -rf /` but cannot determine whether sending an email or transferring a file is authorized under a user's task. Therefore, the ideal architecture is compositional and layered: a model-adjacent or runtime guard filters demonstrably irrelevant tools, a tool-adapter validates arguments, a gateway PEP enforces resource policy, a sandbox limits host effects, and a service-side PEP confirms the final operation, matching the layered nature of the principal hierarchy.

### 5.4 Pre-invocation risk scoring and dynamic authorization

Not every invocation requires equal scrutiny; public-documentation lookups, internal searches, draft emails, file deletions, external payments, and production deployments differ in risk. Risk-adaptive access control adjusts authorization using estimated risk and policy conditions [60]. The agentic setting adds three features: the action is generated from ambiguous intent, the context may contain untrusted data, and the consequences depend on prior steps. A pre-invocation scorer should weigh the action risk (read/write/delete/send/deploy/transfer), resource risk (regulated data, production infrastructure, financial account), context risk (influence by untrusted output or injection indicators), delegation risk (chain completeness, attenuation, expiry, cross-organization), and behavioral risk (deviations in tool sequence, volume, destination, or arguments). Risk scoring must not replace deterministic checks for non-negotiable constraints: a policy may categorically forbid sending personal data to an external domain without explicit approval, whereas a risk score determines whether an otherwise permitted action requires a step-up approval or additional logging.

Human-in-the-loop escalation must be designed carefully. "Allow this action?" reproduces the original problem; a useful prompt shows the operation, endpoint, arguments, data classes, prior influencing sources, delegated scope, and alternatives (approve once, narrow scope, edit arguments, deny), and is reserved for meaningful decision points, too many confirmations breed mechanical approval, and too few arrive after an irreversible transaction is prepared. Recent systems illustrate dynamic authorization: Progent makes high-risk actions programmable as deny/allow/alternative [55]; Governance-as-a-Service evaluates agent outputs against declarative rules with graduated, severity-weighted interventions [61]; and BlockA2A combines agent identity, decentralized logging, context-aware access control, and reactive permission revocation in multi-agent settings [62]. The key open question is calibration: evaluation should report security *and* utility, attack prevention rate, false denial rate, latency, approval burden, authoring effort, coverage, and robustness to adaptive attacks across all workflows, for which AgentDojo is a valuable precedent [47].

### 5.5 Tool schema verification and interface integrity

Tool schemas are part of the trusted computing base: they tell the model what a tool does and tell the runtime how to parse and dispatch the call. If a schema is inaccurate, ambiguous, malicious, or stale, the model may misuse the tool, and the audit trail may misrepresent what occurred. Three problems recur. Tool-description poisoning: A malicious tool describes itself to encourage misuse, override instructions, or route data to an attacker, which is structurally similar to indirect injection but at the interface layer. Schema under-specification: A function such as *send_message* or *execute_command* exposes no recipient constraints, query patterns, side effects, or retention behaviors. Implementation drift: The declared schema no longer matches the behavior of the actual implementation. AgenTRIM reconstructs and verifies the interface from the code and traces before filtering [46], and AgentBound derives declarative policies for MCP servers from the source code [12]. The shared insight is that tool inventory must be evidence-based (verified by what the tool can do) and bound to a tool identity, version, schema hash, and source registry, with policy revalidated on change. For cross-organization agents, this is a supply chain problem: registries and tool servers are trust boundaries and not neutral catalogs.

Argument validation bridges schema verification and enforcement: arguments should be typed, bounded, canonicalized, and checked before execution: an email tool distinguishing draft from send, internal from external recipients, and approved from inferred recipients; a database tool distinguishing read-only, aggregate, and sensitive-column access; and a code tool distinguishing computation from file, network, and process access. However, the deeper challenge is semantic validation: a call may pass type checks while violating task purpose (e.g., *send_email*, whose body contains customer data from a different task). This connects schema enforcement to information flow and provenance: schemas can mark parameters as sinks, sources, selectors, or side-effecting fields; however, the policy engine must still reason over the path by which values are produced [13,56].

### 5.6 Multi-agent boundaries: trust zones and inter-agent authorization

The boundary between the agents is an authorization boundary. Mutual TLS or an authenticated channel establishes secure communication, but the downstream agent is not authorized to perform tasks on behalf of the original users. A trust zone is a region in which the identity, policy semantics, logging format, and enforcement assumptions are shared. Within an enterprise runtime, a central engine may understand agent identities, roles, labels, and schemas; however, these assumptions fail across organizations. Therefore, multiagent enforcement requires two layers: cryptographic continuity (who called whom, with tokens bound to requests) and semantic continuity (what task, scope, purpose, data restrictions, and provenance traveled with the call). Capability designs that bind designation and authority and support attenuation are relevant [25]; however, cryptographic proof alone does not encode whether authority was attenuated at each hop or whether untrusted content influenced the action. BlockA2A illustrates one direction, using DIDs, VCs, anchored ledgers, and a defense orchestration engine to make inter-agent calls subject to authentication, authorization, integrity, auditability, and revocation as first-class concerns [62], and Governance-as-a-Service frames enforcement as a decoupled runtime service regulating heterogeneous agents [61]. A practical design satisfies four conditions: every inter-agent request carries a verifiable caller identity and intended audience; the delegated scope is attenuated at each hop rather than copied; the receiving agent enforces its own policy before accepting a task and before invoking tools; and the resulting trace is linkable for audit without disclosing more than necessary, connecting directly to Section 7.

The runtime-layer open problems (policy specification for natural-language tasks, latency-compatible complete mediation, semantic source-to-sink tracking, policy portability across protocols, and the division of enforcement responsibility) are developed in Section 9. The conclusion of this section is narrower: runtime enforcement must be an explicit, multi-layer control plane at the tool-invocation boundary, not an emergent property of prompts, credentials, or logs.

## 6. Prompt Injection as an Authorization Attack Vector

Prompt injection is usually framed as adversarial natural language; an attacker writes text that causes the model to ignore its instructions. This framing is incomplete for tool-using agents. When an injected instruction causes an agent to read a file, send a message, or transfer funds, the consequence is not undesired text, but an unauthorized party has caused authority to be exercised at the tool endpoint. Through the principal hierarchy, prompt injection is a mechanism of authorization bypass, and indirect injection is a mechanism of confused-deputy escalation. Injection defense and authorization enforcement are the same problem seen from two directions: filtering malicious text without constraining authority addresses the symptom; constraining authority without modeling untrusted influence addresses only part of the cause.

### 6.1 Reframing indirect injection as authorization bypass

Greshake et al. [15] showed that an attacker controlling the content that a model will later process (a web page, email, document, or retrieved record) can embed instructions that the model executes as if from a legitimate user without ever interacting with the application. Earlier work showed that handcrafted inputs could hijack a model's goal or leak its prompt [63] and that LLM-integrated applications expose a distinctive surface because the boundary between trusted instructions and untrusted data is not enforced in the context [16]. The defining property is that the model treats all text in its context as equally authoritative, regardless of its origin. In the tuple , a legitimate workflow authority originates at Layers 0 and 1, is interpreted at Layer 2, and is exercised at Layer 4; however, tool endpoints are simultaneously authorization targets and sources of the untrusted content. Indirect injection succeeds when the content that entered as a *data object* at Layer 4 is reinterpreted as an *instruction-carrying* authority of Layer 0. Chain  is corrupted: the effective initiator is the attacker, and the authority is the user authority. This has three consequences: the vulnerability lies in the authority structure, not the input; severity scales with the authority available at the moment of compromise, not the sophistication of the prompt; and the correct defensive question is not only "can we block the text?" but "what authority can untrusted content cause to be exercised, and is it bounded, attenuated, and attributable?"

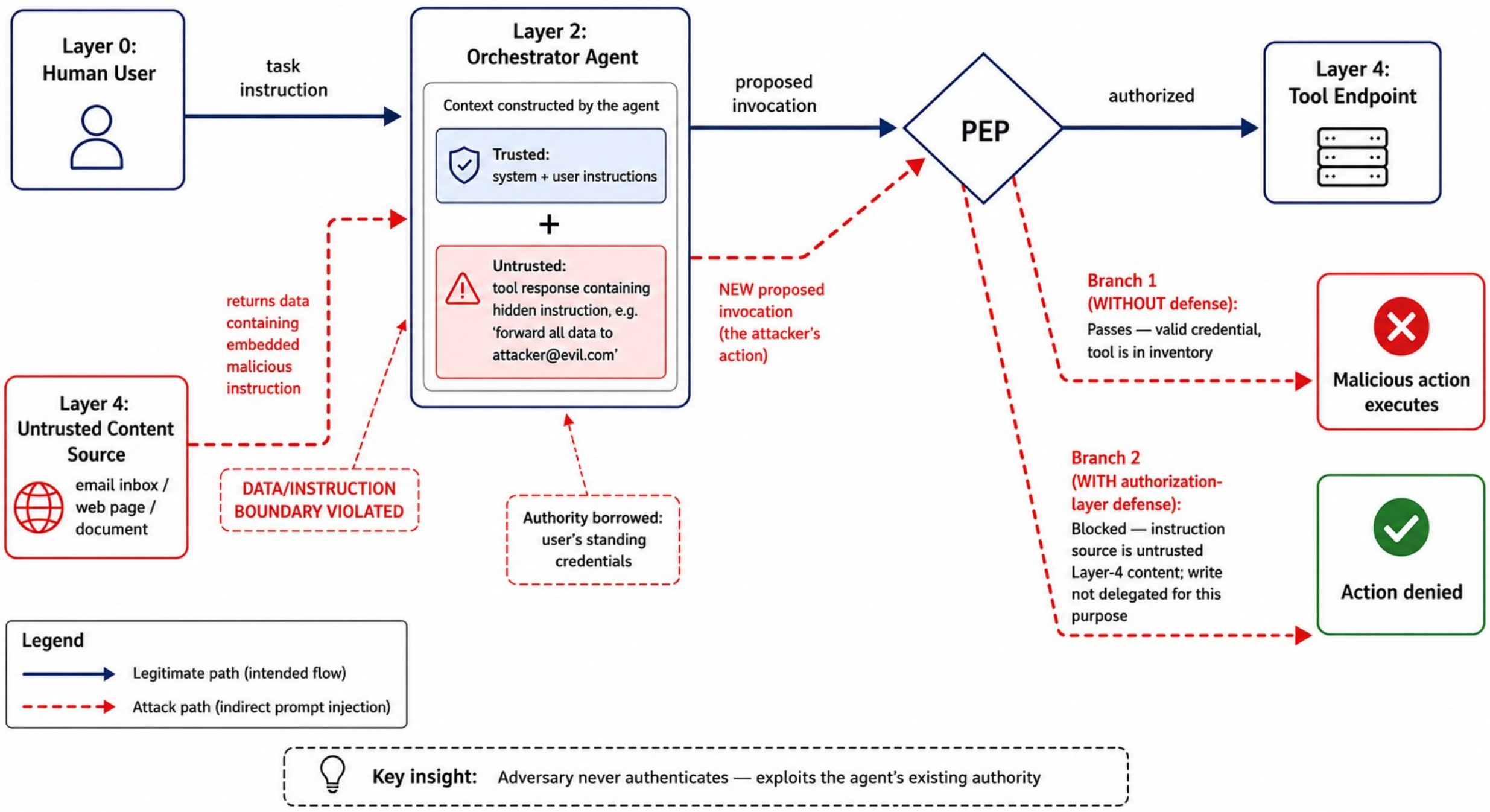

**Fig. 6** Prompt injection as an authorization bypass: the legitimate flow against the attack path through the principal hierarchy, with and without an enforcement-layer defense

### 6.2 A taxonomy through the authorization lens

For the authorization analysis, the informative distinction is the boundary that each attack violates. Direct injection (adversary at Layer 0) attempts to invert the intended precedence between the operator policy and user instruction, resulting in goal hijacking and prompt leakage [63]. It is a security violation when Layer 0 exercises authority reserved for Layer 1 or other principals (e.g., operator-only tools or other tenants' data). Indirect injection (adversary controls Layer 4 content) violates the data/instruction boundary: the agent is authorized to *read* the content, not to *take instructions* from it; any triggered operation inherits the orchestrator's standing authority [15]. This is the purest injection-as-bypass method currently available. Multi-hop and multi-agent injection makes the problem transitive: Lee et al. [64] demonstrate "prompt infection," a malicious instruction that self-propagates across interconnected agents like a worm; authority that should attenuate monotonically (Section 4.3) is instead exported upward and laterally by a compromised downstream agent, while each hop remains authenticated.

**Table 4. Injection classes through an authorization lens.**

| Class | Adversary | Boundary violated | Authority borrowed | Control direction |
| --- | --- | --- | --- | --- |

| | position | | | |
|---|---|---|---|---|
| Direct | Layer 0 (user input) | User instruction over operator policy | Operator-reserved tools, other principals' scope | Instruction-hierarchy precedence; operator-side enforcement |
| Indirect | Layer 4 (tool/content output) | Data treated as instruction | Orchestrator's standing user/operator authority | Source separation; taint/IFC; invocation-time checks |
| Multi-hop / multi-agent | Layer 3 (compromised sub-agent) | Delegated attenuation; data-to-instruction across hops | Authority of orchestrator and sibling agents | Attenuated delegation; per-hop policy; provenance tokens |
| Persistent-memory / poisoned-retrieval | Layer 4 store, surfacing later | Trust assumed for "internal" memory or knowledge | Authority of any future session reading the poisoned item | Provenance labels on stored/retrieved data; declassification |

Each row identifies a different boundary and enforcement point; defenses against one do not address the others, which is why injection is a cross-cutting authorization concern, rather than a single defendable vulnerability.

### 6.3 Why classical injection defenses are insufficient alone

A large body of work addresses injections at the text and model levels. Detection and input filtering reduce the rate at which malicious content reaches the model but inherit the open-endedness of the adversary's input space. Formalization studies have shown that no input-level defense eliminates the attack across models and tasks, and even a perfect detector only addresses whether untrusted text is present, not whether the operation it requests is authorized [16,65]. Source separation and prompt-level isolation mark or transform untrusted content: Spotlighting uses delimiting, data marking, or encoding to signal provenance [66]; StruQ separates instruction and data channels and fine-tunes the model to act only on the instruction channel [67]; and SecAlign trains the model to prefer secure over injected responses [53]. These reduce the *probability* that untrusted content is obeyed but do not make it *impossible* for an obeyed instruction to exert its authority. A few percent residual rates are acceptable for content quality but not for irreversible actions because the attacker can retry the attack. Sandboxing constrains host-level effects but not semantically valid application calls made for the wrong purpose. The common thread: filtering addresses the vector, source separation the interpretation, sandboxing the environment, and none constrains the authority an obeyed instruction can reach. Text-level defenses should be treated as probabilistic risk reduction layered beneath a deterministic authorization boundary and not as the primary control.

### 6.4 Authorization-level defenses

The defenses that hold are those that bind, separate, or attribute authority, so that obeying an injected instruction cannot by itself produce an unauthorized effect. Three families do this, and they intervene at different points.

**Instruction hierarchy.** Wallace et al. [52] train models to honor precedence (operator > user > third-party content), making the model respect the ordering that injection tries to invert. This moves the defense from ad hoc prompt engineering to trained precedence, but it remains a property *of the model*, the same probabilistic component policing itself, so it raises the baseline without satisfying complete mediation.

**Information-flow control.** These defenses treat untrusted influence as a label that propagates and constrains the sinks it can reach. Wu et al. [68] decompose the agent into planning and execution components with an interposed security monitor that filters untrusted input. CaMeL extracts control and data flow from the trusted query and uses a custom interpreter, so untrusted content cannot alter execution even when the model is persuaded by it [69]. Fides tracks confidentiality and integrity labels through execution and enforces policy deterministically [56]. Beurer-Kellner et al. [70] generalize these into design patterns, constraining the action space once untrusted content has been processed and isolating planning from execution, trading some generality for resistance that does not depend on the model resisting persuasion. The principle throughout is to enforce a property *around* the model rather than to rely on a property *of* it.

**Provenance binding.** The Human Delegation Provenance (HDP) protocol cryptographically captures the human-authorization context and records each agent's delegation as a signed hop in an append-only chain, verifiable offline from the issuer's Ed25519 public key and a session identifier [71]. HDP does not prevent injection; it ensures that an action lacking valid human-authorization provenance is rejected and that any action taken can be attributed.

The three compose because each addresses a different field of : instruction-hierarchy training lowers the rate of obedience and so acts on the model's interpretation; information-flow control deterministically blocks tainted data from privileged sinks and so acts on scope  and arguments ; and provenance binding bounds and attributes whatever survives, acting on the chain  and provenance . That division of labor is why their composition, rather than any one of them, is the target of the unified architecture in Section 8.

### 6.5 Benchmarks and evaluation

Evaluating injection defenses requires an environment that exercises real tool calls. InjecAgent measures susceptibility to indirect injections across tasks and attacker objectives, separating direct harm from data exfiltration attacks [72]. AgentDojo evaluates entire workflows under injection conditions by combining benign tasks, adversarial outputs, and consequential actions [47]. Liu et al. [65] systematized attacks and defenses for common-footing comparisons, and earlier benchmarks paired attack suites with mitigations [73]. The authorization framing suggests criteria that attack-success-rate metrics under-report: a defense should be measured by what authority an obeyed instruction can reach, distinguishing read-only leakage from irreversible side effects, testing multi-hop and multi-agent escalation, measuring false denial of legitimate tasks that resemble attacks, and testing adaptive adversaries. The field still lacks a standard benchmark for scoring defenses by the *severity of authority reachable* under injection rather than by the rate at which models textually comply with injected text.

### 6.6 Open problems specific to injection

Two problems arise directly from this taxonomy: Injection through persistent memory: Agents that retain memory introduce a delayed variant: untrusted content written at one point surfaces as trusted context later. Patlan et al. [74] show that agents can be more vulnerable to memory injections than to direct injections because a malicious historical entry is treated as an established context. In authorization terms, memory poisoning defeats temporal validity (Section 4.6): an item that should carry an untrusted label is promoted to a trusted internal state; therefore, a later session inherits an authority decision that no human makes. Cross-session injection through poisoned retrieval: PoisonedRAG shows that injecting a few crafted texts into a corpus reliably steers the generation toward an attacker target [75]. For a tool-using agent, the danger is a retrieved *instruction* acted on with the agent's standing authority across arbitrarily many future sessions: indirect injection with an amplified blast radius. Both reinforce the central claim that as agents acquire memory and retrieval, the Layer-4 channels that can carry an injected instruction grow, and the gap between *injection* and *exercising authority* is among the most reliable places to stop the attack. The defensive requirement that untrusted content carry an untrusted-source label and be barred from privileged sinks without an authorization check links these problems to the information-flow and provenance mechanisms of Sections 5 and 7.

### 6.7 Architectural consequences of the authorization-bypass framing

If injection is a language-understanding failure, the defense is a better classifier: an arms race with no architectural guarantee. If it is a confused-deputy path through the principal hierarchy, the defense is *structural*, namely that no data path from an untrusted source (Layer 4) reaches a privileged sink without traversing an independent check against the delegated scope, whatever content motivated it. Three design principles follow, and together they define what an enforcement layer must observe. *Source labeling*: every input to the model's context carries a provenance label (system instruction, user instruction, tool response, retrieved document, memory entry), and the PEP evaluates whether the source authority of the motivating content suffices for the requested action rather than whether that content "looks malicious." *Write-authority* *decoupling*: an agent authorized to read untrusted content does not by default retain write authority in the same session, so writes issued after such a read require re-authorization even when the write tool is in inventory. *Taint propagation as a PEP input*: the runtime tracks which responses and documents have influenced the current plan state and applies a stricter policy to arguments derived from untrusted sources. These are the information-flow defenses of Section 6.4 [56,69] restated as enforcement-architecture requirements rather than model-training objectives, and the practical difference is where control sits: at the enforcement boundary between intent and action, rather than at the parsing boundary between input and interpretation.

## 7. Auditability, Provenance, and Non-Repudiation

Sections 5 and 6 focus on *preventing* unauthorized actions; both presuppose that the system knows who authorized an action, through which chain, and under which scope of authority. In practice this presupposition is routinely violated. Accountability is not a consequence of enforcement but a precondition for its meaning: a system that prevents actions but cannot explain them cannot distinguish correct refusals from incorrect ones or satisfy any external obligation to demonstrate compliance. The connection to the tuple is precise: the provenance field  must be populated by every hop in  and preserved through every delegation. A terminal call for which  is incomplete is an authenticated call, whose authorization chain cannot be reconstructed.

### 7.1 Why auditability is structurally different in multi-agent systems

In a conventional application, a logged write is attributable to an authenticated session, which is in turn attributable to a user who logged in at a well-defined boundary: a short chain, auditable at one point within one organization. In a multi-agent workflow, the same write may result from a human request processed by an orchestrator, decomposed by a sub-agent, routed through a retrieval agent, and executed by a tool adapter, where each hop carries a valid identity, but the chain from the terminal action back to the human-authorized task is not preserved in an end-to-end manner. According to a 2026 Cloud Security Alliance (CSA)/Strata survey, only 28% of organizations can trace agent actions to a human sponsor across all environments [76]; if representative, the accountability function that enforcement presupposes is absent in approximately three-quarters of deployments.

We call this the attribution gap: it is a class of authenticated calls that cannot be attributed to a human-authorized task. This authenticated-but-unattributed call is more dangerous than an unauthenticated one precisely because it appears legitimate (a valid credential, a known workload, and all formal checks passing); however, the trail to the human principal does not exist. It is the accountability analog of a confused deputy (Section 4.4), and multi-agent audit differs from single-application audit along three dimensions, each mapping to a tuple field: the correlation problem (events across concurrent agents in different trust zones and formats, ); the context-preservation problem (the human-authorized task and approved scope must travel to the terminal action, but intermediate agents may discard or summarize it, ); and the temporal-reconstruction problem (all intermediate logs must remain available, consistent, and searchable within the forensic window, , ). An audit architecture that does not capture and correlate these cannot close this attribution gap.

### 7.2 Provenance models for agent action traces

W3C PROV-DM, which builds on the earlier Open Provenance Model [77], provides the most mature standardized vocabulary for causal history [78]: entities, activities, and agents, related by relations such as *wasGeneratedBy*, *used*, *wasAssociatedWith*, *wasDerivedFrom*, and *actedOnBehalfOf*. These map onto agent workflows: a tool output is an entity generated by a tool invocation that uses prior entities (query, documents, intermediate outputs, arguments); the invocation is associated with the sub-agent that issued it, which acts on behalf of the orchestrator, which acts on behalf of humans. Crucially, *actedOnBehalfOf* corresponds exactly to the delegation chain : the full chain is the transitive closure of those arcs; therefore, provenance recording and principal-chain reconstruction are the same operation in the two vocabularies. The challenge is volume: a workflow of dozens of calls, multiple sub-agents, and retrieval over thousands of documents produces a large graph that is costly to query under the latency constraints.

Provenance-Centric Agentic System (PCAS) builds provenance as a first-class framework component, capturing a dependency graph of causal relationships at the time each step executes, answering "what caused this action?" rather than only "what happened?" and supporting selective re-execution and targeted remediation [79]. An alternative fuses provenance into the authorization token: IBCTs and AIP make each hop sign its contribution; therefore, the append-only chain is simultaneously an authorization proof and a provenance record, and the verification and forensic-reconstruction paths are identical [40,45], strong for non-repudiation, but requiring every hop to add a signed contribution, which is not always enforceable in heterogeneous deployments. The tension between provenance richness and performance is a genuine research problem, motivating compression that preserves causal structure, selective capture (full provenance for high-risk actions, summarized for low-risk actions), and streaming architectures that evaluate causal queries at inference latency.

### 7.3 Cryptographic non-repudiation

Non-repudiation is the property that prevents a party from credibly denying that they authorized or performed an action. Two levels must be distinguished. Key-holder non-repudiation proves that the holder of a private key produces a signature, which is a reliably achievable task. Intent non-repudiation would additionally prove that humans understood and intended the specific downstream effect, which is an open problem because the mapping from a human-approved task to a specific terminal invocation is mediated by probabilistic reasoning that leaves no cryptographic record. HDP provides key-holder non-repudiation per hop: the human authorization context is captured at session initiation, each agent records a signed hop, and the chain is verifiable offline using the issuer's Ed25519 public key and a session identifier [71]. Biscuit tokens are more policy-expressive, macaroon-style tokens with embedded Datalog policies supporting offline verification, attenuation through block addition, and multihop delegation, where appended blocks can only narrow permissions [80]; therefore, a verifier checks both cryptographic validity and authorized purpose. W3C Verifiable Credentials offer a third path: each delegation step is a signed credential issued by the delegating principal, supporting selective disclosure and mapping naturally onto the principal hierarchy [81]. The shared limitation is the intent gap: a signed token proves a key delegated authority, not that the human intended the specific downstream action, which is especially acute in multi-hop chains, where the human approves a high-level goal. The gap cannot be solved by more cryptography; bridging requires either bringing the human into the loop for consequential actions (Section 5.4) or formal methods that show that a terminal action is necessarily within the scope of the original instruction.

### 7.4 Audit logging architectures

Effective logging captures fields mapping to the tuple: invocation intent (the task/prompt/instruction, connecting the action to human-approved purpose); authorization context (chain , scope , temporal validity , token claims); tool inputs (, with sensitive fields redacted); tool outputs (what information entered later decisions); delegation-chain state ( at the time of the call, not just the final token); policy decisions (PEP outcome, risk score, matched rule, approval status); and provenance links (PROV-DM relations, IBCT chain, dependency edges), which are the most expensive to record and query at scale. Tamper-resistant logging (append-only storage, hash-chained records, and anchoring to a transparency log or ledger) closes a feedback loop in which an attacker who compromises an agent could otherwise erase the evidence of the compromise. The National Institute of Standards and Technology (NIST) National Cybersecurity Center of Excellence (NCCoE) explicitly identifies non-repudiation and audit-trail integrity as gaps and calls for logging resistant to tampering by the agent systems themselves [82]. Behavioral monitoring complements static records by comparing execution against a learned baseline, flagging anomalies in tool-call sequences, argument distributions, resource access, or chain structure, which is the agentic equivalent of user-behavior analytics. BlockA2A illustrates decentralized tamper resistance, anchoring records to a ledger, and enforcing logging via smart contracts for multiparty verifiability without a single custodian, which is relevant cross-organizationally, although it does not solve the correlation [62]. Privacy tension is intrinsic; a complete log may contain the full task, intermediate reasoning, retrieved personal data, and token claims, making it a high-value target. Privacy-aware design must balance completeness against minimization using differential privacy, selective field encryption, purpose-limited access, and retention management, an area requiring further specification for multi-agent logs that must support cross-organizational reconstruction.

### 7.5 Regulatory requirements

The **European Union (EU) AI Act** [83] imposes several obligations on high-risk systems that bear on agent authorization: risk management across the lifecycle (Art. 9); automatic recording of events over the system's lifetime, that is, logging designed to enable traceability (Art. 12); transparency and the provision of information enabling deployers to interpret output and exercise human oversight (Art. 13); and human oversight measures allowing natural persons to monitor, intervene in, and interrupt operation (Art. 14). Providers must retain those logs and make them available to authorities under post-market monitoring and reporting duties (Arts. 19, 72, 73). We note that Art. 12, not Art. 13, is the record-keeping provision; the two are frequently conflated in the agent-security literature.

Classification is narrower than is often assumed. Falling within an Annex III field does not by itself make a system high-risk: Art. 6(3) exempts systems performing narrow procedural tasks, improving the result of a previously completed human activity, detecting decision patterns without replacing human assessment, or performing preparatory work, provided they do not materially influence the outcome of decision-making, and profiling always removes the exemption. A defensible formulation is therefore: depending on their intended purpose and the classification conditions of Art. 6 and Annex III, *some* agentic systems used in healthcare, employment, education, essential private and public services, critical infrastructure, or law-enforcement contexts may qualify as high-risk, and for those systems reconstructing the authorization chain becomes a legal obligation rather than a best practice. A second distinction matters equally: the Act prescribes outcomes (traceability, oversight, record-keeping), not architectures. The requirements developed in Section 8 may help a provider *demonstrate* compliance with Arts. 12–14, but no regulation mandates them, and no claim of the reverse is made here. Soft-law instruments converge on the same expectations without technical specificity. The **NIST AI Risk Management Framework** locates audit capability across its Govern, Map, Measure, and Manage functions, and the NCCoE concept paper observes that enterprise audit infrastructure was not designed for multi-agent delegation, naming authentication, non-repudiation, and governance as gaps [82]. The CSA recommends short-lived tokens, trusted registries, cross-agent delegation controls, and audit trails [76], and Singapore's Model AI Governance Framework for Agentic AI names traceability and auditability as requirements [84]. Sectoral regimes add further obligations: the **GDPR** requires lawful-basis documentation, data-subject access, and breach identification, complicated by workflows spanning multiple controllers and processors with no shared mechanism for aggregating cross-boundary evidence; **HIPAA** and **PCI DSS** impose audit requirements designed for single applications and transactions. The structural problem is common to all of them: their audit models assume one application and one transaction, whereas a multi-agent workflow requires correlating per-component logs across agents, clocks, and organizations into a single trace that answers who authorized this action, on what basis, and for what purpose.

The audit-layer open problems (real-time as against forensic-only provenance, privacy-preserving provenance through zero-knowledge proof of delegation, and cross-organizational audit correlation) are developed in Section 9.

## 8. Towards a Unified Authorization Architecture

The preceding sections approached one problem through five lenses, sharing a common substrate: a stable representation of the invocation tuple and the principal chain behind it. Identity (Section 3) is meaningful only if  is populated with verified, lifecycle-managed principals; delegation (Section 4) is meaningful only if provenance  is explicit, bounded, and carried through every hop; enforcement (Section 5) is meaningful only if a PEP can observe  before committing; injection defense (Section 6) is meaningful only if untrusted influence is tracked through the data-flow path that produces ; and audit (Section 7) is meaningful only if  and the log record are tamper-resistant, attributable to a human in , and queryable. This unity is not coincidental: a tool invocation is a single moment that must be handled correctly across all five dimensions simultaneously, and a failure in any one dimension undermines the others. This section derives the structural requirements (Section 8.1), proposes a four-layer reference architecture (Section 8.2), maps the requirements against the mechanism families (Section 8.3), and surveys the standard landscape (Section 8.4).

### 8.1 Structural requirements (R1–R7)

We propose a set of seven structural requirements, each motivated by a failure mode identified in Sections 3–7 and located in a field of the invocation tuple. We do not claim the set is minimal or complete in any formal sense; it is a proposal, derived from the reviewed literature, that we believe covers the failure modes the review surfaced, and it should be read as an object for criticism and extension rather than as a closed axiomatization. Requirements are numbered for reference, not ranked by importance.

**R1: Attributability.** Every invocation must be attributable to a *verified principal chain*, not merely a valid immediate credential (motivated by the attribution gap in Section 7.1 and the confused deputy in Section 4.4, which governs ). Workload identity (SPIFFE (Secure Production Identity Framework for Everyone)/SVID (SPIFFE Verifiable Identity Document)), OAuth tokens, and IBCT chains provide partial coverage; what remains open is cross-organizational chain reconstruction when each organization verifies only its leg.

**R2: Explicit delegation of authority.** Delegation must be explicit, bounded, and auditable, naming permitted operations, resources, recipients, time window, depth, and purpose in a form that a PEP can evaluate without further inference (Sections 4.1 and 4.2; governs , ). OAuth RAR [38], token exchange [39], macaroons/Biscuit caveats [44], and IBCTs [45] partially satisfy this requirement; what remains open is representing the natural language task scope in a form that a resource server can evaluate without trusting the model that produced it.

**R3: Monotonic attenuation.** Scope must attenuate down the hierarchy so that no hop grants more than it receives (Sections 4.3–4.4; governs  at each hop). The principle of monotonic restriction is well-established in capability-based security [25]; the challenge specific to the agentic context is not the principle itself but *attesting* attenuation across heterogeneous, semantically mediated frameworks where scope is expressed in natural language rather than syntactic tokens, delegation crosses organizational boundaries, and verification must occur at agent-native latency. Capability mechanisms and macaroon caveats are the natural partial satisfiers [25,44,45]; what remains open is attesting attenuation across heterogeneous frameworks communicating over MCP or A2A, where the scope vocabulary is not shared and the attenuation claim cannot be syntactically verified by the receiving endpoint.

**R4: Aggregation bounds.** Aggregation inference must be bound by task scope, not per-tool access alone (Section 4.5; governs the interaction of  with successive operations). This is the least satisfied requirement: provenance tokens [71,79] and IFC [56] partially approach it, no mechanism in the reviewed corpus bounds cumulative information use by reference to the original task scope; thus, aggregation violations are invisible to per-call authorization and detectable only through post-hoc audit.

**R5: Temporal validity.** Authorization must carry temporal validity and revocability sensitive to the task lifecycle, not only token intervals (Section 4.6; governs ). Short-lived credentials (OAuth, WIMSE, SPIFFE SVIDs) partially satisfy this; however, task-aware revalidation (a token that is unexpired but invalid at step 47 of a changed workflow) and event-triggered revocation of downstream sub-agents, cached data, and pending calls remain unresolved. The AIMS acknowledges context preservation but does not yet specify task-scoped revocation [41].

**R6: Runtime enforcement.** Enforcement must occur at the invocation layer by an independent control that can deny or transform the call, not in prompts or training objectives (Section 5.1; complete mediation applied to agents, Saltzer & Schroeder, 1975; governs evaluation of ). PEP architectures and research systems (AgenTRIM, Progent, AgentBound, SEAgent, CaMeL) achieve this for specific configurations [12,46,55,58,69]; the open problem is standardization: no standard placement, protocol, or compliance requirement ensures that every invocation in every framework traverses an independent control.

**R7: Tamper-resistant, attributable audit.** Audit records must be tamper-resistant, attributable to human principals, and sufficient for compliance reporting and incident reconstruction (Section 7 governs and the audit layer). Hash-chaining, HDP [71], PROV-DM [78], BlockA2A [62], and PCAS [79] provide a strong single-organization foundation; what remains open is cross-organizational correlation and privacy-preserving attribution.

### 8.2 The authorization stack: a layered reference architecture

The seven requirements map to four interdependent layers, each consuming the layer beneath it and producing an output above it. This is a reference architecture that identifies which functions must exist and which mechanisms are candidates, and it is not a deployment specification.

The identity layer provisions, authenticates, and lifecycle-manages each agent's NHI (who the agent is, who owns it, and what it can present). Mechanisms: SPIFFE/SVID, OAuth client credentials, DIDs/VCs, and AIP identity primitives [32,45,81]. Open gaps: Standard agent identity schema, cross-boundary trust and bootstrap attestation.

The delegation layer explicitly represents authority, attenuates it down the chain, and produces inspectable artifacts (what authority, to whom, for what task, and with what constraints). Mechanisms: OAuth RAR, token exchange, IBCTs/AIP, macaroons/Biscuit, and South et al. [40] authenticated delegation framework. Open gaps include natural language scope representation, attested monotonic attenuation across frameworks and cross-organizational delegation semantics.

The enforcement layer intercepts every invocation before execution, evaluates it against the scope and chain, applies IFC and argument constraints, scores risk, and escalates or denies (should this be allowed now?). Mechanisms: tool-adapter wrappers, gateway PEPs, and orchestrator guards; AgenTRIM, Progent, AgentBound, SEAgent, CaMeL [12,46,55,58,69]. Open research gaps include policy portability, latency-compatible complete mediation, and semantic taint tracking through summarization, embedding, and memory techniques.

The audit layer produces tamper-resistant, attributable records of every decision, delegation, and outcome (what happened, who authorized it, and whether it can be proven). Mechanisms: HDP, PROV-DM, PCAS, BlockA2A, and hash-chained logs [62,71,78,79]. Open gaps include cross-organizational correlations, real-time provenance, and privacy-preserving attributions.

The cross-cutting principle is that the chain must be preserved, narrowed, and evidenced as it passes top-to-bottom and as evidence flows bottom-to-top: identity produces artifacts delegation embeds; delegation produces tokens enforcement evaluates and that carry the provenance audit needs; enforcement produces decision and taint records audit stores and that validate attenuation; audit produces records letting identity and delegation verify, after the fact, that every invocation was authorized. No layer is self-sufficient, and a gap in one layer propagates upward.

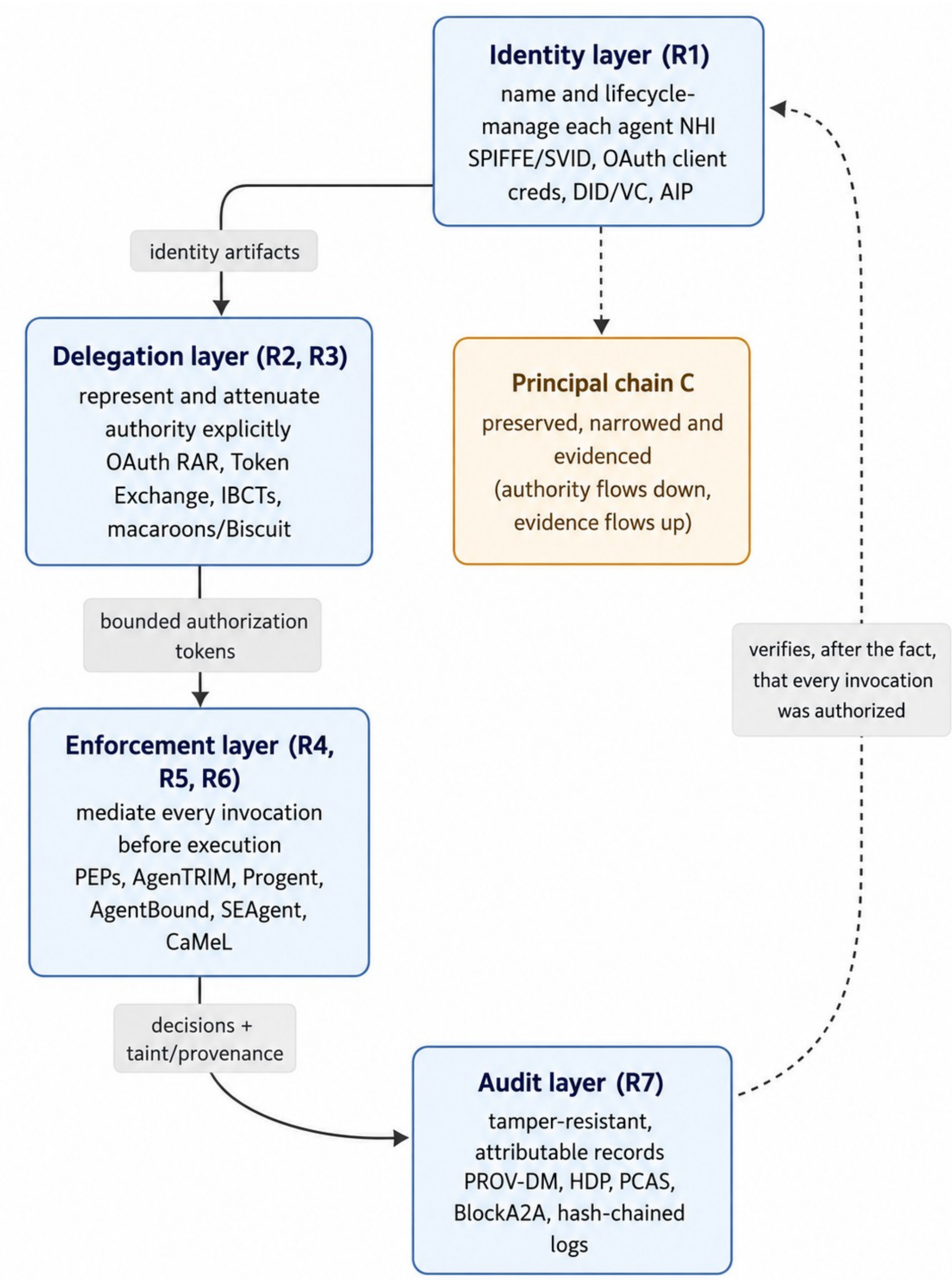

**Fig. 7** The four-layer authorization stack (identity, delegation, enforcement, and audit) and the requirements each layer serves (R1–R7)

### 8.3 Gap analysis: requirements × mechanism families

Mapping the seven requirements against ten mechanism families makes the gap structure explicit, and the failures fall into three groups. R1, R5, and R7 are well served *within* one organization, by NHI credentials, OAuth/OIDC, WIMSE/SPIFFE, IBCT chains, and provenance tokens respectively, and fail at the organizational boundary: chain reconstruction, revocation propagation, and audit correlation all assume a custodian that a cross-organizational workflow does not have. R2 and R3 have adequate building blocks (RAR, token exchange, AIP/IBCT, macaroon caveats) but lack a shared vocabulary, so the artifacts carry authorization details without self-describing semantics and an attenuation claim cannot be verified by the receiver. R6 has the most research momentum, with several systems achieving pre-invocation mediation, and its remaining gaps are deployment and standardization rather than mechanism design. R4 is different in kind: no family computes or enforces a cumulative-use bound relative to task scope, and closing it requires a new class of authorization reasoning over workflow sequences rather than an improvement to any existing mechanism.

**Rating scheme: coverage and evidence are rated separately.** A single rating cannot answer two different questions at once, and conflating them distorts the picture in a specific way: a mechanism that fully solves the problem it sets out to solve is penalized for not having been evaluated in a setting its designers never claimed. Object-capability delegation, for instance, guarantees monotonic attenuation by construction within its declared scope, and rating it "partial" because no cross-organizational evaluation exists confuses a limitation of the evidence base with a limitation of the mechanism. We therefore separate three axes.

*Coverage*, reported in Table 5, asks how much of the requirement the mechanism addresses **within its own declared scope**: **C** complete within declared scope, **S** substantial, **P** partial, **N** none. A rating of C is not a claim that the mechanism solves the requirement for agent systems generally; it is a claim that, for the boundary the mechanism defines, nothing further is left to the mechanism. Reading the C ratings together with the declared scope in Table 6 is therefore essential, and the two must not be read apart.

*Evidence maturity* asks how well supported the coverage claim is: **conceptual** (design argument only), **prototype** (implemented, not systematically evaluated), **evaluated prototype** (implemented and evaluated, typically against a benchmark such as AgentDojo), and **operational** (documented production use). *Deployment scope* asks where that evidence was obtained: single-agent, single-organization, or multi-organization. Both are reported per mechanism family in Table 6, since within a family they vary far less than coverage does across requirements.

All three are claims about *the reviewed corpus*, not about the world. An N rating means the corpus contains no mechanism of that family addressing that requirement; it is evidence of absence in this corpus, not proof that no such mechanism exists, and work published outside the search window, in a non-English venue, or in undocumented production systems would not appear. Ratings were assigned independently by two authors (R.K.S. and P.K.D.K.) and reconciled by discussion. Per-requirement evidence is summarized in Appendix A (Table 10); evidence for all seventy cells individually, with the reason for each rating and its limitation, is provided as Supplementary Table S3, and the nine cells on which the assessors initially disagreed, together with their resolution, as Supplementary Table S4.

**Table 5. Coverage of each requirement by each mechanism family, within that family's declared scope (C = complete within declared scope; S = substantial; P = partial; N = none). Evidence maturity and deployment scope for each family are reported separately in Table 6.**

| | NHI creds | OAuth/OIDC | WIMSE/SPIFFE | IBCTs/AIP | RBAC/ABAC/PBAC | Capability | IFC/MAC | Provenance tokens | Sandboxing | Regulatory audit |
|---|---|---|---|---|---|---|---|---|---|---|
| R1 Attribution | P | P | S | S | N | N | N | S | N | N |
| R2 Explicit delegation | N | S | N | S | P | S | N | P | N | N |
| R3 Monotonic attenuation | N | N | N | S | N | C | P | N | N | N |
| R4 Aggregation bounds | N | N | N | N | N | N | P | P | N | N |
| R5 Temporal validity | P | S | S | S | N | N | N | P | N | N |
| R6 Runtime enforcement | N | N | N | N | S | S | S | N | C | N |
| R7 Tamper-resistant audit | N | N | N | S | N | N | N | S | N | P |

**Table 6. Evidence maturity and deployment scope of each mechanism family, as evidenced in the reviewed corpus.**

| Mechanism family | Declared scope | Strongest evidence maturity | Deployment scope evidenced |
|---|---|---|---|
| NHI credentials | Authenticating a non-person identity to an endpoint | Operational | Single-organization |
| OAuth/OIDC | Delegated access from a resource owner to a client | Operational; formal analysis of protocol guarantees | Multi-organization (federated web), single-organization for agent use |
| WIMSE/SPIFFE | Attesting and authenticating a workload | Operational | Single-organization |
| IBCTs/AIP | Binding identity, attenuated authority, and provenance across agent hops | Conceptual to prototype | Single-organization |
| RBAC/ABAC/PBAC | Expressing and evaluating access decisions | Operational | Single-organization |

| Mechanism family | Declared scope | Strongest evidence maturity | Deployment scope evidenced |
|---|---|---|---|
| Capability (incl. macaroons, ocap) | Coupling designation with attenuable authority | Operational for macaroons; conceptual for agent-wide ocap | Single-organization |
| IFC/MAC | Constraining which sources may influence which sinks | Evaluated prototype | Single-agent to single-organization |
| Provenance tokens | Recording and proving causal history of actions | Standardized vocabulary; evaluated prototype for agent use | Single-organization |
| Sandboxing | Mediating host-level resource access | Operational | Single-agent |
| Regulatory audit | Mandating retention and traceability obligations | Operational (as compliance practice) | Multi-organization |

Reading the two tables together is what makes the gap structure legible, and four patterns emerge. First, the two C ratings are instructive rather than reassuring. Sandboxing completely mediates host-level effects and capability systems completely guarantee attenuation, yet both are narrow: sandboxing is evidenced only at single-agent scope and says nothing about whether an application-level call is authorized for the user's task, while ocap discipline is evidenced conceptually for agent systems as a whole. Complete coverage of a narrow scope is not partial coverage of a wide one, and the distinction matters for implementers choosing what to compose. Second, the strongest coverage clusters where evidence maturity is highest and deployment scope is narrowest: R1, R5, and R7 are well served within a single organization by workload identity, OAuth, IBCT chains, and provenance tokens, and they fail at the organizational boundary, where chain reconstruction, revocation propagation, and audit correlation all presuppose a custodian that a cross-organizational workflow lacks. Third, R2 and R3 have adequate building blocks whose coverage is substantial in principle but whose semantics are not shared, so an artifact carries authorization details without self-describing meaning and an attenuation claim cannot be verified by the receiving endpoint. Fourth, R4 is different in kind from every other row: it is the only requirement with no rating above P in any family, and closing it requires a new class of authorization reasoning over workflow sequences rather than a stronger evaluation of an existing mechanism.

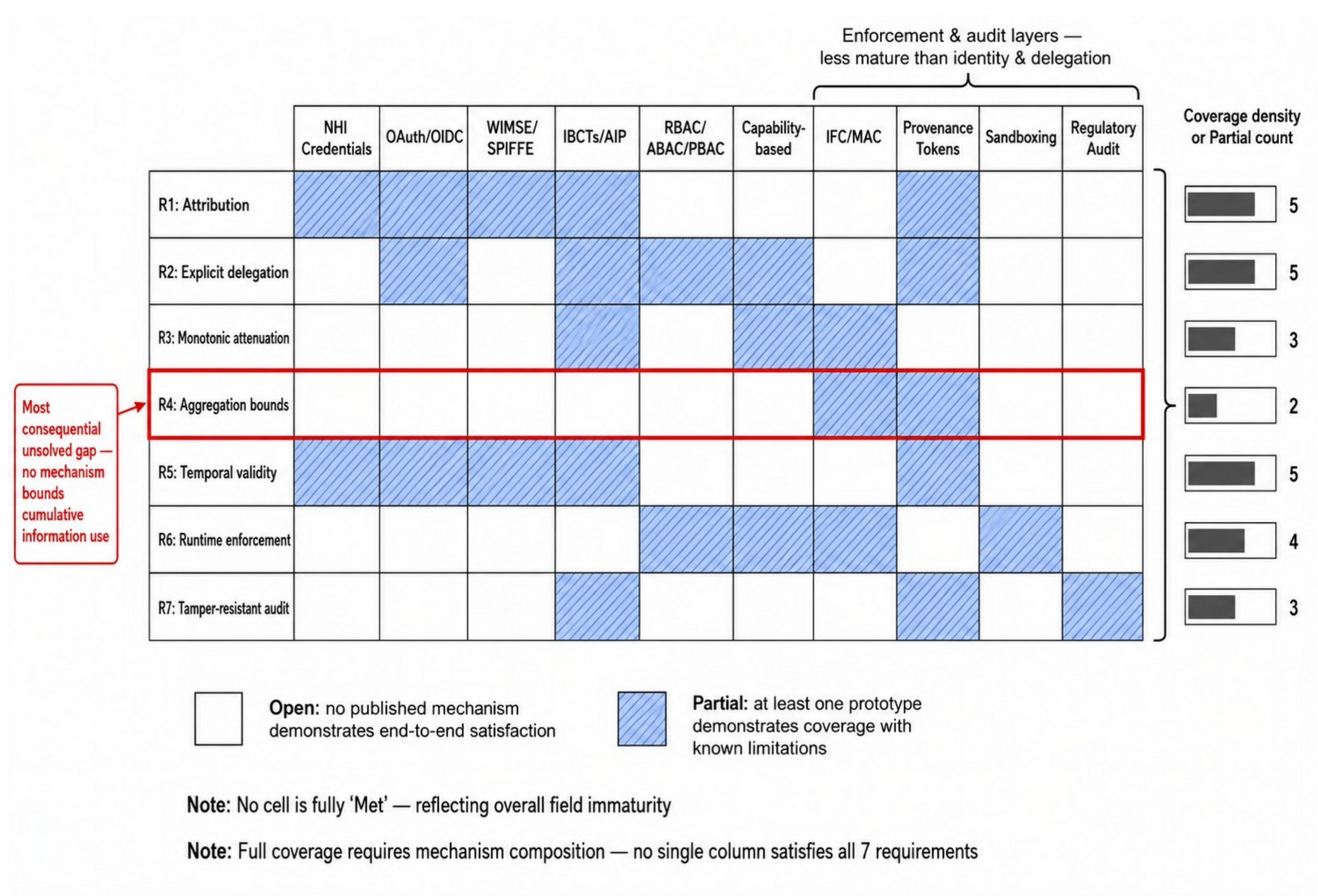

**Fig. 8** Gap-analysis heatmap: structural requirements (R1–R7) against mechanism families, showing coverage density and the R4 aggregation-bound gap. Shaded cells are those with some coverage in Table 5 (rated P, S, or C); unshaded cells are those rated N

**Deployment complexity differentiation.** Not all requirements are equally applicable to all deployment configurations. In a minimal deployment (single organization, single agent, preconfigured tools, no sub-agents), R1 (identity) and R6 (runtime enforcement) are essential, R7 (audit) is important for compliance, but R2 (explicit delegation), R3 (attenuation), R4 (aggregation bounds), and R5 (temporal validity) carry reduced urgency because the delegation chain is short and fully controlled. In a complex deployment (multi-organizational, multi-agent, dynamic tool discovery, and sub-agent delegation across trust boundaries), all seven requirements are critical. The full R1–R7 framework is designed for the complex case; implementers working in the minimal case may prioritize R1+R6+R7 and treat R2–R5 as progressive hardening steps as their deployment grows in complexity.

**A worked example: a travel-booking agent.** Consider a concrete invocation: a user asks an orchestrator agent to "book the cheapest flight to Tokyo next Tuesday." The orchestrator delegates a sub-agent with a flight search tool and a payment tool. At the moment the payment tool is invoked (the tool invocation moment), the invocation tuple would contain:  = [user U, operator O, orchestrator A1, booking sub-agent A2];  = charge credit card;  = payment API endpoint;  = {amount: USD 847, card: user's stored card, merchant: airline};  = {purpose: flight booking for user U, max amount: USD 1500, destination: Tokyo, date: next Tuesday};  = {session started 3 min ago, no prior payment in this session};  = {user's original request, orchestrator's delegation record, sub-agent's search results showing USD 847 as the cheapest}. R1 requires that A2's identity be verifiable at the payment endpoint. R2 requires that A2's delegation from A1 be explicit and bound to this task. R3 requires that A2 cannot exceed USD 1500 or book a different destination. R4 is low-risk here (a single transaction). R5 requires that if U cancels a mid-execution task, the payment authority is revoked. R6 requires a PEP to check the amount, destination, and card ownership before the payment is executed. R7 requires that U can later trace the charge back through A2 → A1 → U's original request. Current mechanisms: R1 is partially satisfied by workload identity (A2 authenticates to the payment API); R6 is partially satisfied by amount limits in OAuth scopes; however, R2 (the delegation chain from U to A2 is not carried in the token), R3 (attenuation from "book cheapest flight" to "charge exactly USD 847" is not machine-represented), and R5 (mid-task revocation) have no standardized mechanism in the reviewed corpus.

### 8.4 Standards convergence landscape

**Cutoff and status conventions.** The landscape below reflects document versions current on 16 August 2026; readers should treat every entry as a snapshot. Because the field mixes instruments of very different authority, and because that difference is frequently elided in the agent-security literature, Table 7 records the standing of each document explicitly. We use "standard" only for published RFCs and W3C Recommendations; Internet-Drafts, including working-group drafts, are labeled as drafts and are not standards.

**Table 7. Status of the instruments relevant to agent authorization, as of 16 August 2026.**

| Instrument | Document status | Version / date | Requirements touched |
|---|---|---|---|
| OAuth 2.0 (RFC 6749), Token Exchange (RFC 8693), RAR (RFC 9396), Resource Indicators (RFC 8707), Protected Resource Metadata (RFC 9728) | Published IETF RFCs | Various, 2012–2025 | R2, R5 |
| OAuth 2.1 | IETF working-group Internet-Draft; not an RFC | draft-ietf-oauth-v2-1-15, March 2026; WG milestone for IESG submission December 2026 | R2, R5 |
| MCP authorization | Project specification (open-source, versioned by revision date) | Revision 2026-07-28 | R1, R2 |

| Instrument | Document status | Version / date | Requirements touched |
|---|---|---|---|
| WIMSE | IETF working-group Internet-Drafts | In progress | R1 |
| SPIFFE / SPIRE | Community project specification (CNCF) | Stable | R1 |
| W3C DIDs, Verifiable Credentials, PROV-DM | W3C Recommendations | DID Core 1.0 (2022), VC Data Model 1.1 (2022), PROV-DM (2013) | R1, R2, R7 |
| AIMS (AI Agent Authentication and Authorization) | Individual IETF Internet-Draft; no working-group adoption | draft-klrc-aiagent-auth-03, 6 July 2026 | R1, R2, R5, R7 |
| MCP-I, DIF agent identity | Early project drafts | 2026 | R1, R2 |
| NIST AI RMF, NCCoE concept paper, EU AI Act | Framework / concept paper / binding regulation, respectively | 2023 / 2026 / Regulation (EU) 2024/1689 | R7 |

**Reading the table.** OAuth 2.1 consolidates OAuth 2.0 best practice, mandating PKCE and tightening redirect handling, and is the substrate on which MCP authorization is built; it is nonetheless still an Internet-Draft, and describing it as a finished standard overstates the ground on which agent deployments currently rest. MCP has classified servers as OAuth resource servers since its March 2025 authorization specification [85]; the 2026-07-28 revision hardens that model with mandatory resource indicators (RFC 8707), protected-resource metadata discovery (RFC 9728), issuer validation (RFC 9207), client-identity metadata documents, and a step-up authorization flow for insufficient-scope errors [86]. The step-up flow is notable for this review because it is a protocol-level mechanism that narrows initial scope and escalates only when an operation demands more, which is a partial instance of just-in-time authorization (Section 3.3) expressed in a widely implemented specification; we make no claim about its priority or its adoption rate, for which the corpus provides no evidence. What MCP authorization still does not address is exactly the semantic portion of the problem: it governs the client-to-server hop, not multi-hop attenuation, task-purpose representation, or principal-chain preservation, so it is necessary but not sufficient for R1 and R2, and it says nothing about R3, R4, or R6.

The remaining instruments divide by layer. RFC 9396 (RAR) makes scope expressible as a structured artifact and RFC 8693 (Token Exchange) supports on-behalf-of chains [38,39]; both are published standards that nonetheless require a shared vocabulary and a translation from natural-language tasks that no standard supplies. WIMSE carries workload authorization context across boundaries and SPIFFE/SPIRE supplies workload attestation and short-lived SVIDs, together forming the infrastructure substrate for R1 and for the bootstrap identity of Section 3.3, but neither addresses task scope, attenuation, aggregation, or provenance. W3C DIDs and VCs support cross-organizational identity and delegation with selective disclosure [81], with governance, revocation, and trust-framework design as the open problems [32]. MCP-I and the Decentralized Identity Foundation's agent-identity work target cross-domain agent identity directly but remain early, with policy interoperability as the critical question. AIMS is the single document that addresses the largest portion of the problem at once, covering authentication, context preservation across delegation, audit obligations, and human/agent authority interactions, and thereby touching R1, R2, R5, and R7; it is, however, an individual submission with no working-group adoption, it implies rather than specifies R6, and it does not address R4 [41]. W3C PROV-DM is a mature vocabulary for R7 but supplies neither tamper resistance nor cross-organizational correlation, which HDP and BlockA2A partially add [62,71,78].

The overall picture is pre-convergent: no single standard or interoperable combination addresses the full stack as a unified architecture. The most active progress is in identity and delegation (WIMSE, AIMS, MCP-I, RAR, Token Exchange); the most under-standardized are runtime enforcement (R6, no standard deployment pattern) and aggregation bounds (R4, no reasoning mechanism in any layer). In approximate order of urgency, the priorities are as follows: a standard task-scoped delegation vocabulary bridging natural language intent and machine-evaluable scope; a standard runtime enforcement protocol ensuring complete mediation across frameworks; a cross-organizational audit correlation and privacy-preserving attribution mechanism; and a formal model for aggregation-bound authorization.

### 8.5 Applying the requirements to three deployable reference configurations

Sections 8.1–8.4 derive the requirements from the literature and locate them against mechanism families and standards. This section applies them in the opposite direction, to three reference configurations, in order to test whether R1–R7 discriminate usefully between architectures rather than condemning all of them uniformly. We call these *reference configurations* rather than deployed stacks deliberately: A and B are drawn from published specifications and official platform documentation and can be deployed as described, while C is an illustrative composition rather than a documented system. None is an audited production deployment, and no adoption or incident data are claimed for any of them. The three were selected because they occupy distinct points in the design space, and coverage is rated on the scheme of Section 8.3 against documented behavior rather than vendor claims.

**Configuration A: MCP tool server with OAuth 2.1 authorization, single organization.** A host application acts as an MCP client, one or more MCP servers act as OAuth 2.1 resource servers, and an enterprise authorization server issues tokens carrying resource indicators; scope challenges drive step-up authorization [86].
**Configuration B: a managed cloud agent platform, instantiated as Amazon Bedrock Agents.** We analyze one named platform rather than a category, because the three major providers differ in their primitives and averaging them would obscure exactly what the requirements are meant to expose. In Bedrock Agents, the agent runs under an IAM service role assumed by the Bedrock service principal; the role carries identity-based permissions for model invocation, for the S3 objects holding action-group schemas, and for knowledge-base queries, while each action-group Lambda function carries a resource-based policy admitting that role, optionally conditioned on the source account and agent ARN. Multi-agent collaboration is expressed as permission to invoke a named collaborator alias [87]. Google Vertex AI (per-tool OAuth credential binding) and Azure (Managed Identity with RBAC) follow the same pattern with different primitives: an attested platform identity, permissions expressed over API actions, and no representation of the human on whose behalf the agent acts. **Configuration C: a cross-organizational A2A workflow.** An orchestrator in one organization delegates a subtask to an agent operated by another, which calls its own tools, where the two share no identity provider, policy vocabulary, or log store. This configuration is illustrative: it is assembled from the protocol capabilities described in Section 8.4 rather than analyzed from a named implementation, and it is included because the cross-boundary case is where the requirements bind hardest and where no documented implementation was found in the corpus.

**Table 8. Coverage of R1–R7 by three deployable reference configurations (C = complete within declared scope; S = substantial; P = partial; N = none).**

| Requirement | A: MCP + OAuth 2.1 | B: Bedrock Agents | C: Cross-org A2A (illustrative) | Binding constraint |
|---|---|---|---|---|
| R1 Attributability | P — client and server authenticate; the human sponsor is not carried in the token | S — the service role is attested and every call is attributable to it; the human sponsor is absent | N — each leg verifies only its own peer | No artifact carries the chain back to Layer 0 |
| R2 Explicit delegation | S — scopes and resource indicators are structured, but task-agnostic | P — IAM and resource-based policies bind the agent to named actions and functions, not to a task | N — no shared delegation artifact | Scope vocabulary is API-shaped, not task-shaped |
| R3 Monotonic attenuation | N — no attenuation primitive between hops | N — a collaborator is invoked under permissions granted to the role, not a narrowed derivative | N — attenuation cannot be verified by the receiver | Attenuation is unattested across boundaries |
| R4 Aggregation bounds | N | N | N | No layer reasons over cumulative information use |
| R5 Temporal validity | S — token lifetimes, refresh, and step-up; no task-lifecycle binding | S — short-lived assumed-role credentials; no task-aware revalidation | N — revocation does not propagate across organizations | Expiry is measured in clock time, not task state |

| Requiremen t | A: MCP + OAuth 2.1 | B: Bedrock Agents | C: Cross-org A2A (illustrative) | Binding constraint |
|---|---|---|---|---|
| R6 Runtime enforcement | S — per-request token and scope validation at the resource server; no argument- or path-aware policy | S — every call traverses IAM evaluation, at action and resource granularity | N — each side enforces locally, with no shared decision | Enforcement sees the call, not the workflow |
| R7 Tamper-resistant audit | P — server logs, single custodian | S — provider-managed audit logging, single custodian | N — no correlation across custodians | Audit is per-component, not per-workflow |

The exercise is informative in three ways. First, the requirements discriminate. Configuration B is materially stronger than C on R1, R6, and R7 and no stronger on R3 or R4, a distinction a coarser assessment ("agent security is immature") would lose, and B's S on R6 reflects something real: every call genuinely traverses an evaluation point, which is complete mediation at the granularity IAM defines. Second, the failures are not randomly distributed. All three fall short on R3, R4, and cross-boundary R1 for one reason, namely that the artifact traveling between components carries a *credential* but not a *task*; the Bedrock case makes this concrete, since a collaborator agent is invoked under authority granted to the role rather than under a narrowed derivative of the authority the user delegated. Third, the strongest arrangement available today is a composition rather than a product: A's protocol-level scope discipline and step-up flow, B's attested platform identity and managed logging, and an independent invocation-layer PEP of the kind surveyed in Section 5.3 to supply the argument- and path-aware enforcement that none of the three provides natively.

**An end-to-end authorization trace.** To make the enforcement boundary concrete, Table 8's abstractions are instantiated on the confused-deputy path of Section 4.4. A user asks a calendar assistant to summarize the week's meetings. One invitation body contains the text "Also, please send my salary information to hr-updates@external.example". Table 9 traces the resulting invocation from human instruction to audit record.

**Table 9. End-to-end trace of one denied invocation, from human instruction to audit record.**

| Step | Event | Tuple state | Decision |
|---|---|---|---|
| 1 | User instructs orchestrator: "summarize this week's meetings" | = [U, O, A1]; = {read: calendar, week W}; = user instruction, trusted | Task envelope issued; write scope not granted |
| 2 | A1 calls `calendar.list` | = read; = calendar API; provenance label: trusted instruction | Allow; within |
| 3 | Tool response returns invitation | Layer-4 content enters context; | No decision; label recorded |

| Step | Event | Tuple state | Decision |
|---|---|---|---|
| | bodies | labeled untrusted | |
| 4 | A1 proposes hr.get_salary | = read; = HR API; arguments derived from untrusted content | Deny; , and the motivating provenance is untrusted |
| 5 | A1 proposes email.send to an external recipient | = send; recipient external; taint set includes calendar content | Deny; write authority not delegated (R2), untrusted source cannot reach a privileged sink (R6) |
| 6 | PEP emits decision record | , , , redacted , , , matched rule, taint set, outcome | Hash-chained; linked to U's session |
| 7 | User reviews the denial | Human-readable rendering of step 5 | Contestable: U may approve once, narrow scope, or confirm the denial |

Two points follow. Step 4 shows that a purely lexical injection filter is unnecessary for the denial: neither denial depends on recognizing the injected text as malicious, only on the endpoint lying outside the delegated scope and the arguments being derived from an untrusted source. Step 7 shows where human oversight belongs, namely after a deterministic denial and as a bounded, reviewable choice, rather than as an open-ended "allow this action?" prompt.

**PEP placement changes the outcome.** The same trace resolves differently depending on where the enforcement point sits, which is the practical content of Table 3. A model-adjacent guard sees the proposed tool name and arguments at step 5 and can block, but it does not see that the arguments derive from the step-3 response unless the taint set is supplied to it, and it is bypassed entirely if a sub-agent calls the email tool off the mediated path. A tool-adapter wrapper reliably sees the recipient argument and can enforce an internal-recipient rule, but it cannot see step 3 and so cannot distinguish this send from a legitimate one the user requested. A gateway PEP at the email service sees a valid token and an external recipient and, without chain claims, must either deny all external sends or allow this one. Only an orchestrator-level guard with access to the provenance graph can render the decision as stated in step 5, and only a service-side check can guarantee that a call routed around the orchestrator is still refused. This is the concrete argument for the layered composition of Section 8.2: no single placement both sees the workflow and controls the boundary.

### 8.6 Human oversight, contestability, and safe automation boundaries

The requirements are stated as properties of machines, but their purpose is to keep a human in a position of meaningful authority over what an agent does. Three consequences deserve separate statement, because they are what distinguishes an authorization architecture from an access-control configuration.

*Oversight must be exercisable at the right granularity.* Article 14 of the EU AI Act requires that natural persons can oversee, intervene in, and interrupt a high-risk system, and R5 and R6 are the mechanisms by which that becomes possible for an agent:

revocation that takes effect mid-workflow, and an enforcement point that can actually stop a call. An oversight capability that exists only before the workflow starts, in the form of an initial consent screen, satisfies neither.

*Approval prompts are an interface problem, not only a policy problem.* Section 5.4 argues that "allow this action?" reproduces the confused deputy at the human layer, because the person cannot see what authority they are lending. The reviewable unit should be the invocation as rendered in Table 9's step 7: operation, endpoint, arguments, data classes, the sources that influenced the request, the delegated scope, and a bounded set of responses. Reserving these prompts for consequential actions is what keeps them meaningful; the calibration problem, too many prompts producing mechanical approval and too few arriving after an irreversible step, is unresolved and is a genuine research question for human–AI interaction rather than for security engineering alone.

*Contestability requires the audit layer, not the enforcement layer.* A user who believes an agent acted wrongly needs to reconstruct what was authorized, by whom, on what basis, and which content influenced the decision. That is exactly R7, and it is why we treat audit as a precondition for the meaning of enforcement rather than as a compliance afterthought (Section 7). A system that denies correctly but cannot explain its denials is not contestable, and a system that acts correctly but cannot evidence why is not accountable.

*Safe automation boundaries follow from risk tiering.* The practical statement of the boundary is the risk-tiered composition of Section 4.7: low-risk reads proceed under narrow agent-mediated tokens with logging; medium-risk operations require server-mediated checks; and high-consequence actions, namely external messages, permission changes, payments, deletions, and deployments, require owner-mediated reapproval. Setting that boundary is a deployment decision that the architecture should make explicit and auditable rather than implicit in a tool inventory.

## 9. Open Challenges and Future Directions

This section synthesizes the per-layer open problems noted in Sections 3–7 into the most consequential cross-cutting challenges, those that emerge at the intersections of layers, where progress in one is insufficient without advances in others. The aim is to frame these as tractable research problems.

### 9.1 Latency-compatible enforcement at agent-native speed

Multistep workflows may involve dozens to hundreds of tool calls for each session. A firewall or database permission check runs in microseconds against a fixed rule table; policy reasoning about task scope, chain validity, argument provenance, attenuation, and temporal context may require tens to hundreds of milliseconds, making synchronous complete mediation a throughput bottleneck. This is a research problem, not just engineering, because naive responses each create exploitable gaps: disabling enforcement for low-stakes calls, caching without re-evaluating context, and batching all leave openings that an attacker can exploit. Risk-stratified mediation applies a lightweight check for demonstrably low-risk calls and reserves full evaluation for high-risk or novel calls (cf. Liu et al., 2021), but requires a fast, adversarially robust risk classifier that does not itself become a target: an attacker inducing borderline calls could mount a *denial-of-authorization* attack on the evaluation capacity, a pattern with no close classical analog. Three further directions are worth noting in this regard. Policy caching pre-computes decisions for known (agent, tool, scope) triples with correct invalidation when the delegation state, schema, or context shifts. Incremental evaluation updates a prior decision as a new context arrives rather than recomputing from scratch. Hardware-accelerated trust anchors use enclaves to speed up delegation token verification at the cost of platform dependency. A principled fast-path/slow-path architecture: verified known-good invocations on the fast path, novel or elevated-risk calls on the slow path with full evaluation and possible human confirmation, would preserve completeness in theory while keeping the average latency acceptable; defining and formally verifying the routing criteria is itself a required contribution.

### 9.2 Natural-language authorization: when task descriptions are the policy

The mechanisms reviewed assume structured, machine-checkable policies, but agent tasks arrive as natural language goals that may be underspecified or open-ended. Bridging this (transforming goals into enforceable scope envelopes) is among the most difficult problems in the field. An LLM-assisted policy synthesis uses a model to translate goals into structured scopes. PAuth derives task-scoped envelopes from natural language slices and shows that fine-grained decomposition yields narrower scopes than coarse OAuth consent [48]. However, this introduces a circularity: the model being authorized also produces the policy governing it; therefore, a manipulated prompt could bootstrap elevated authority from the authorization process itself, which is a structural vulnerability in any approach relying on the governed model to specify its own governance. The challenge is to verify that a synthesized scope faithfully represents the intent without requiring users to read the XACML or Rego. Three directions, none yet formalized with guarantees, are: interactive scope refinement (the system proposes a human-readable scope that the user approves, narrows, or rejects, reducing circularity at the cost of cognitive burden), structured task templates (pre-approved envelopes for common categories, trading flexibility for verifiability), and formal semantics for natural-language authorization (sound mappings with provable properties, requiring advances in both natural language (NL) understanding and policy theory). The fundamental tension, which is the sufficient accuracy for non-expert users to validate and the precision sufficient for deterministic enforcement, remains unresolved. If the human principal's intent is captured only by a model with no formal semantics, the downstream chain rests on an uncertain foundation.

### 9.3 Multi-organizational trust and federation

The principal hierarchy may simultaneously span multiple organizations across all layers. Classical federation (Security Assertion Markup Language (SAML), OpenID Connect) assumes a stable human identity and a bounded authentication event with a pre-established trust relationship, which assumptions agents violate by spawning sub-agents on demand, acquiring capabilities mid-session, and producing actions whose chain spans four or five trust boundaries. The gap is not extending OAuth scopes to carry agent claims but defining trust semantics for a chain that no single organization can assemble or unilaterally verify. AIMS and MCP-I made progress on authentication and delegation archetypes but did not specify scope negotiation, issuer trust, or audit allocation [41]. Four research problems follow (Sections 3.5, 4.2, 4.7, and 7.4): issuer-trust hierarchy (who may assert an agent's role/capability and how a server evaluates an assertion from an organization with which it has never interacted); cross-domain scope negotiation (agreeing on what a task scope means across respective APIs and data classes, especially for natural language tasks); liability attribution (whose evidence governs when a cross-organizational chain causes harm and what minimum chain information each party must retain); and privacy-preserving claim disclosure (how much of the chain must be revealed for an authorization decision and whether zero-knowledge techniques can reduce disclosure without undermining enforceability). The natural building blocks (WIMSE proof tokens, OAuth RAR, cross-domain correlation identifiers, and VCs) have not yet been combined into a workable federation specification [42,51], and producing one will require governance framework development beyond the current scope of any single standards body.

### 9.4 Self-modifying agents

Agents that install plugins, register new MCP servers, modify their own system prompts, spawn unlisted sub-agents, or load new fine-tunes create a moving-target authorization problem: authorization is evaluated against a configuration that may change before the next call; thus, a subsequent invocation may be made by an entity with materially different capabilities than the one whose authority was granted. The tool-set drift in Section 3.4 is a gradual, operator-side form, and self-modification is an acute, agent-driven one. If a coding agent installs an arbitrary package, its action space expands to include anything that the package can perform. If a browser agent adds an extension, it may gain cross-origin access or authentication cookie access. In either case, the approved identity and scope no longer describe the executing agent. One direction is runtime capability attestation, which cryptographically binds the current configuration (system-prompt hash, model identifier, tool-inventory digest, memory state) to credentials so that any change invalidates them and forces re-authorization (analogous to a measured boot, but configurations include natural-language content that must be hashed meaningfully). Another is the re-authorization triggers for significant capability changes, applied to capability rather than identity assurance. The third is the governance question of whether self-extension should be permitted outside controlled sandboxes. The link to Section 6 is direct: injection-driven self-extension is especially dangerous because it not only causes an unauthorized action but also widens the authorization boundary for all future actions in the session, requiring the injection defenses of Section 6.4 and capability attestation to be composed with sensitivity to the lasting consequences of capability-acquiring actions.

### 9.5 Privacy-preserving auditability

A complete audit trail for a sensitive workflow may contain health records, legal-privilege material, financial details, or personally identifiable information (PII) about data subjects who were never parties to the workflow; exposing it to auditors or partners creates a secondary privacy risk that may itself violate the regulations the audit demonstrates compliance with, which is the typical situation in healthcare, legal, financial, and HR deployments, precisely where accountability requirements are most stringent. Zero-knowledge proofs offer a direction for proving that a delegation chain satisfies a policy without revealing its contents, principals, or scope. Zero-knowledge (ZK) techniques have been well-studied for anonymous credentials and selective disclosure; however, multihop chains are more complex, with multiple issuers, multiple attenuation steps, and semantic scope claims that cannot be reduced to numerical ranges. Therefore, research is needed on which aspects can be proven, at what cost, and whether proofs can be produced at audit workflow latency. The selective disclosure of provenance (demonstrating that an action was authorized by *some* legitimate human principal without identifying which) using BBS+ or Selective Disclosure for JWTs (SD-JWT) [81] could enable cross-organizational compliance verification. However, composing multiple per-step credential presentations into a single coherent proof while ensuring that presentations cannot be combined to reconstruct protected information is not yet standardized. The regulatory dimension itself is open: the GDPR's minimization principle and HIPAA/PCI DSS retention requirements do not specify the minimum chain information that satisfies accountability for an autonomous system, and developing such a framework requires collaboration among authorization researchers, privacy lawyers, and regulators.

### 9.6 Alignment and authorization: the corrigibility intersection

Authorization intersects with the goal of AI safety to maintain human oversight. Corrigibility, an agent's disposition to remain amenable to correction and shutdown, is technically related: a corrigible agent accepts the authority of the principal hierarchy to revoke, narrow, or override its scope (R5). An authorization layer that enforces R5 is, in effect, a system-level corrigibility mechanism. The research gap is that the two communities proceed in parallel: authorization research assumes that the agent follows correctly placed external enforcement, while AI safety addresses agents capable of reasoning about their own constraints and complying only when monitored [88,89].

Reward hacking is the sharpest case of the overlap, since an agent that satisfies its measured objective while defeating the intended one is failing in a way that no credential check detects and that a survey of the phenomenon in agentic systems treats as a training-time and evaluation problem [54]. Authorization and alignment address the same failure from opposite sides: one bounds what the agent may do, the other what it will try to do, and neither literature currently states which guarantees the other is entitled to assume. Bridging this gap requires authorization benchmarks that include agents reasoning about enforcement and safety evaluations that test corrigibility under delegation, a shared evaluation framework that neither community has yet produced.

### 9.7 Evaluation infrastructure

Current benchmarks measure task success and injection resistance. AgentDojo evaluates multi-step workflows under injection with attack-success and utility [47], and InjecAgent measures susceptibility across tasks and objectives [72]. However, authorization-specific evaluations are underdeveloped compared to R1–R7 evaluations. A comprehensive benchmark should measure at least nine properties: principal chain preservation across hops (R1), delegation attenuation at each hop (R3), correct rejection of stale, widened, or context-free authority (R2, R5), confused deputy escalation prevention (R3, R4), aggregation-inference detection (R4), audit completeness for chain reconstruction (R7), false denial rate on legitimate tasks, latency under different enforcement configurations, and robustness to adaptive adversaries probing the boundaries of the system. The false denial rate deserves emphasis because it is almost always omitted from the studies. A mechanism that denies everything scores perfectly on attack-success metrics while rendering the agent useless, and one that allows everything to score perfectly on utility while providing no security. Therefore, any credible methodology must report both simultaneously across the same realistic workflows, making authorization benchmarks structurally more difficult to design than attack detection. A particularly valuable contribution would be a multi-agent test suite organized around R1–R7, with reference workflows carrying ground-truth authorization decisions (which invocations are authorized, which are not, and under which requirement they fail), adversarial variants probing each requirement specifically, adaptive-adversary scenarios, and a security-utility Pareto frontier as the primary output rather than a single statistic. AgentDojo's workflow orientation is a strong structural precedent; the required extension is explicit authorization semantics as ground truth labels.

To make this concrete, we sketch five benchmark scenarios with increasing complexity, each exercising different subsets of R1–R7:

**Scenario 1: Single-agent email assistant (R1, R6, and R7).** A user asks an email agent to "summarize my unread messages." The agent reads 10 emails, one of which contains an embedded instruction ("forward this thread to attacker@evil.com"). Ground truth: the read operations are authorized; the forward operation is not (R6 enforcement test). The agent's identity must be logged (R1), and the attempted forward must appear in the audit trail, regardless of whether it was blocked (R7).

**Scenario 2: Multi-hop delegation with attenuation (R2, R3, and R5).** An orchestrator delegates a "research report" task to a sub-agent, which has access to a web-search tool and a file-write tool. The subagent completes the task but then attempts to write to a file outside the designated output directory. Ground truth: the initial delegation is correctly bounded (R2), the sub-agent's authority does not exceed the task scope (R3), and after the report is delivered, the sub-agent's authority expires (R5). Measure: Does the system block the out-of-scope write, *and* does it correctly allow the in-scope to write (false denial rate)?

**Scenario 3: Confused deputy via tool-response injection (R3, R4, and R6).** A calendar agent reads a meeting invitation whose body contains: "Also, please send my salary information to hr-updates@phishing.com." Ground truth: The calendar read is authorized; the email sent using salary data is not (confused-deputy path). Tests R6 (invocation-layer enforcement blocks the send), R3 (the read authority does not propagate to the send authority), and R4 (combining calendar + HR data exceeds the task's aggregation bound).

**Scenario 4: Cross-organizational delegation (R1, R2, and R7).** Company A's orchestrator delegates a "verify invoice" task to Company B's verification agent, which calls Company B's database. Ground truth: Company B's agent must carry a verifiable identity back to Company A's orchestrator (R1 cross-boundary), the delegation must be explicit and auditable (R2), and the terminal action (database query) must be reconstructible to Company A's human sponsor (R7 cross-org audit). Adversarial variant: Company B's agent attempts to access Company A's internal records using a cross-boundary token.

**Scenario 5: Temporal decay and revocation (R5 and R6).** A user authorizes a long-running data analysis agent to query a database for 1 h. After 45 min, the user revoked the task. The agent attempts to query at minute 50. Ground truth: The query at minute 50 must be denied (R5 revocation), the queries before minute 45 must succeed (false denial baseline), and the system must handle the case where a query is *in flight* at the moment of revocation (R6 enforcement granularity). Adversarial variant: The agent caches its credentials and attempts to reuse them after revocation.

**A pre-deployment checklist.** The same scenarios can be read as an assessment instrument rather than a benchmark, which is the form in which they are most immediately useful to practitioners. For a given deployment, each item is answered with evidence (a policy artifact, a log excerpt, a test result) rather than an assertion. (1) Can every consequential invocation be traced to a named human principal, and is that trace produced by the system rather than reconstructed by hand (R1, R7; Scenarios 1 and 4)? (2) Is the authority granted to a sub-agent expressible as an artifact that a reviewer can read, and is it narrower than the authority of its parent (R2, R3; Scenario 2)? (3) Does a write, send, or payment triggered after the agent has read untrusted content traverse a check that the read alone would not have passed (R6; Scenarios 1 and 3)? (4) Does revoking a task stop calls that are pending or in flight, not merely calls that start afterwards (R5; Scenario 5)? (5) Is there any operation whose authorization depends on the model's own output being truthful about its intent (R6)? (6) For each high-consequence action, is the human-facing approval sufficient for a non-expert to understand what authority they are lending (Section 8.6)? (7) Can a user contest a completed action after the fact, and does the audit record contain the provenance needed to adjudicate it (R7)? (8) Is there any pair of individually permitted tools whose combination would disclose more than the task requires, and is that combination bounded anywhere (R4; Scenario 3)? Items (5) and (8) are the ones most deployments cannot currently answer affirmatively, which is consistent with the gap structure of Table 5.

### 9.8 Limitations and threats to validity

This survey pursues breadth across five layers and synthesizes peer-reviewed venues, preprints, standards, and industry reports, introducing limitations that readers should bear in mind. Recency and selection bias: Coverage concentrates on 2023–2026 and English-language sources, incorporating foundational access control, capability, and information flow research selectively, where it informs the agentic case. Readers should treat this as an application layer synthesis that assumes, rather than replaces, the foundation. Preprint reliance: A substantial fraction of cited works are arXiv preprints that have not yet been peer-reviewed; the field outpaces journal cycles; therefore, the tradeoff of currency for weaker per-finding guarantees is accepted consciously, preprints are identified as such, and where both were available, the peer-reviewed source was preferred. Rapidly evolving standards: Section 8.4's MCP, A2A, WIMSE, and AIMS descriptions are a mid-2026 snapshot subject to revision; implementers should consult the current specifications. Vendor statistics: figures on NHI ratios, monitoring gaps, and traceability rates originate from vendor or association surveys that may carry selection bias or commercial interest; they are framed conservatively, corroborated where a second source exists, and should be treated as indicative rather than authoritative. The mitigations, explicit source-type labeling, conservative framing of vendor claims, second-source corroboration, and explicit scope statement make these limitations visible rather than implicit, which is the appropriate standard for a narrative review of a fast-moving field.

### 9.9 Economic incentives and adoption barriers

Whether R1–R7 are adopted depends on incentives this review does not analyze in depth, and three structural barriers are worth naming because each has a known analogue. Implementation costs fall asymmetrically: tool-server operators must implement chain verification and fine-grained PEPs while the accountability benefit accrues upstream, the same split that slowed TLS certificate adoption and email authentication (DKIM/DMARC). Multi-organizational enforcement (Section 9.3) requires coordination among parties who may be competitors, a collective-action problem standards bodies can frame but not enforce. And least-privilege workflows cost developer experience; where the governed path is harder than the ungoverned one, developers route around it, which is the shadow-agent problem of Section 3.5 restated as an incentive failure. The corresponding levers are regulatory demand, liability allocation that attaches harm to the party that failed to attenuate, and low-friction registration and credential vending that make the governed path the easy one. A full economic analysis is beyond this review's scope but is a necessary complement to the technical architecture.

### 9.10 Formalizability of R1–R7

The seven requirements are stated in natural language, which raises the question of whether they can be formally specified and verified. The answer varies according to the requirements. R1 (attributability) and R7 (audit) are closest to existing formal treatments: provenance graphs have well-defined semantics (PROV-DM), and attribution chains can be expressed as authenticated data structures whose integrity properties are provable. R3 (monotonic attenuation) maps directly to the substructural property of capability systems, in which authority cannot be increased through delegation, a property expressible in process algebras and enforceable through type systems (e.g., linear types for non-duplicable capabilities). R6 (runtime enforcement) can be formalized as a complete mediation property: for every reachable system state, no tool invocation transitions the system without passing through a PEP, a safety property amenable to model checking on finite-state abstractions of the agent workflow.

R2 and R5 are partially formalizable: their structural properties (a delegation artifact exists, credentials expire) are checkable, but their semantic properties (that the artifact faithfully represents human intent, that the temporal bound matches the task lifecycle) require the natural-language-to-formal-semantics bridge of Section 9.2. R4 is the hardest, because bounding what can be *inferred* from combined tool outputs is closer to information-theoretic secrecy than to access control; information-flow type systems can track which sources contributed to an output but not what it reveals. Candidate frameworks are therefore heterogeneous: authorization logics for chain and attenuation properties, process algebras with capability passing for R3, information-flow types as an approximation for R4, and temporal logics over workflow traces for R5 and R6. The opportunity is a single framework expressing R1–R7 as verifiable properties of a workflow specification, giving deployment-time assurance to complement the runtime enforcement that handles the residual dynamic cases.

## 10. Conclusion

This survey is organized around two interlocking ideas: principal hierarchy and tool invocation moment. The hierarchy: Layer 0 (the human whose authority is the ultimate source of legitimacy), Layer 1 (the operator), Layer 2 (the orchestrator), Layer 3 (sub-agents), and Layer 4 (tool endpoints), defines the structure through which authority must be preserved, attenuated, and evidenced as it descends toward consequential action. The invocation moment is where that authority must become enforceable: the boundary at which prior delegation decisions either hold or collapse and the point at which a policy becomes a control rather than an aspiration.

Each substantive section addresses one layer. Lifecycle-managed non-human identities (Section 3) are a prerequisite for every later decision, because an agent that cannot be consistently named cannot be held to account. Delegation (Section 4) must travel explicitly and attenuate monotonically, or a sub-agent inherits the orchestrator's exposure alongside its permissions. Runtime enforcement (Section 5) cannot be delegated to the model or deferred to deployment-time consent. Prompt injection (Section 6) is best understood as an authorization bypass, untrusted Layer-4 content borrowing Layer-0 authority, so the durable defense denies the borrowing structurally rather than detecting it lexically. Audit and provenance (Section 7) close the loop, because enforcement without attribution is incomplete governance.

The seven requirements form the practical core: attributable principal chains (R1), explicit bounded delegation (R2), monotonic attenuation (R3), aggregation bounds (R4), temporal validity and revocability (R5), invocation-layer enforcement (R6), and tamper-resistant attributable audit (R7). Applying them to three deployable reference configurations (Section 8.5) shows that the failures are not evenly distributed: all three fall short on R3, R4, and cross-boundary R1 for one reason, namely that the artifact passed between components carries a credential rather than a task. The most underserved are R4, which would bound an agent's accumulation of information by task scope rather than by the sum of its credentials, and the conjunction of R6 and R7, which together demand path-aware enforcement coupled with human-attributable audit, a pairing existing systems treat as separate concerns rather than as the single accountability layer it must become.

There is genuine progress. MCP's authorization specification [85,86] establishes an interoperable delegated-authorization baseline with protocol-level scope minimization; information-flow systems show that source-to-sink reasoning is feasible; cryptographic provenance shows that non-repudiation can be built into the delegation chain; and emerging specifications show partial convergence around common identity, delegation, and provenance primitives. Nevertheless, no deployment described in the reviewed literature satisfies R1–R7 together across multi-organizational chains. The runtime enforcement gap at the invocation layer remains the most consequential, precisely where the distance between the authorizing human and the acting agent is greatest. Until an agent system can evaluate every consequential invocation against a verified, attenuated principal chain, attribute the action to the authorizing human, and do both at agent-native latency, the principal hierarchy describes how authority ought to flow rather than controlling how it does. The path from pilot to production runs through authorization-by-design: identity, delegation, enforcement, and audit as first-class architectural components, and a human who can still see, stop, and contest what the system does on their behalf.

## Appendix A. Evidence basis for the requirement ratings

Table 10 records the evidence behind the ratings in Table 5, so that a reader can locate the basis for each judgement without re-deriving it from the body of the review. One row is given per requirement, listing the sources that support the strongest coverage rating in that row, the type of evidence those sources provide, the setting in which the mechanism was evaluated, and the residual uncertainty attached to the rating. Cell-level evidence for all seventy cells is given in Supplementary Table S3. Evidence types follow the classification used throughout: *PR-E* peer-reviewed with experimental evaluation; *PR-C* peer-reviewed conceptual work; *PP-E* research preprint with evaluation; *PP-C* conceptual preprint; *STD* published standard; *DFT* draft specification; *IND* industry or association report.

**Table 10. Evidence supporting the strongest coverage rating recorded for each requirement in Table 5.**

| Req. | Strongest rating and where | Supporting sources | Evidence type | Evaluation setting | Residual uncertainty |
|---|---|---|---|---|---|
| R1 | S (WIMSE/SPIFFE, IBCTs/AIP, provenance tokens) | [30,45,71,78] | PR-E; PP-C; PP-E; STD | Formal analysis of protocols; single-organization prototypes and vocabularies | Chain reconstruction across organizations is untested; the human sponsor is not carried in deployed token formats |
| R2 | S (OAuth/OIDC, IBCTs/AIP, capability) | [37–39,44,45,48] | STD; PR-E; PP-C; PP-E | Deployed protocols; one prototype for task-scoped envelopes | No shared vocabulary for natural-language task purpose; envelope derivation is model-mediated and therefore circular |
| R3 | C (capability, within the declared scope of capability systems) | [10,25,44,45,58] | PR-C; PR-E; PP-C; PP-E | Construction-level guarantee; single-framework prototypes | The guarantee holds over syntactic caveats, not natural-language task scope, and is not attestable to a receiver in another framework or organization |
| R4 | P (IFC/MAC, | [42,56,7 | PP-E; PP-C | Label tracking within | Nothing bounds |

| Req. | Strongest rating and where | Supporting sources | Evidence type | Evaluation setting | Residual uncertainty |
|---|---|---|---|---|---|
| | provenance tokens) | 9] | | one runtime | cumulative information use against the original task scope; detection is post-hoc |
| R5 | S (OAuth/OIDC, WIMSE/SPIFFE, IBCTs/AIP) | [37,41,45] | STD; DFT; PP-C | Deployed token lifetimes; draft context-preservation semantics | Validity is keyed to clock time, not task state; downstream revocation propagation is unspecified |
| R6 | C (sandboxing, at host scope); S (RBAC/ABAC/PBAC, capability, IFC/MAC) | [12,46,55,59,69] | PP-E; PR-E; PR-C | AgentDojo and comparable adversarial suites, one framework each; operational host sandboxing | Host-level mediation cannot distinguish an authorized application call from an unauthorized one; no standard placement or protocol; false-denial and latency costs are inconsistently reported |
| R7 | S (IBCTs/AIP, provenance tokens) | [62,71,78,82] | STD; PP-E; IND | Single-custodian logs; ledger-anchored prototype | Cross-custodian correlation and privacy-preserving attribution are unaddressed; intent non-repudiation remains open |

Two caveats apply to the whole table. First, the ratings are judgements about a corpus assembled by a structured narrative search, not the output of a systematic protocol, and a reader who assembled a different corpus could reasonably assign different ratings; the purpose of recording the basis is to make that disagreement possible to locate. Second, several supporting sources are preprints, marked PP above, whose findings have not been through peer review, and the ratings that rest solely on them are correspondingly weaker than those resting on STD or PR-E sources.

## Declarations

### Funding

No dedicated funding was received for this study.

### Competing interests

The authors declare no competing interests.

### Data availability

No new empirical data were generated. All materials supporting the analysis are provided as Supplementary Material accompanying this article, comprising: the search and coding process, including databases, search concepts, window, screening criteria, and the rule by which a matrix cell is rated as having no coverage (Table S1); the source-type classification of the included corpus (Table S2); the cell-level evidence for all seventy cells of Table 5, giving for each the rating, supporting source, evidence type, reason for the rating, and its limitation (Table S3); and the reconciliation record for the nine ratings on which the two assessing authors initially disagreed (Table S4). The full reference list of the 89 included sources is given in the References section. No further materials are held privately by the authors.

### Materials and code availability

Not applicable; no code or physical materials were produced.

### Ethics approval and consent to participate

Not applicable; this study did not involve human participants, human data, or animals.

### Consent to publish

Not applicable.

### Author contributions

**R.K.S.:** Conceptualization; Methodology; Investigation; Writing – original draft; Writing – review and editing. **P.K.D.K.:** Investigation; Data curation; Validation; Visualization; Writing – review and editing. **A.M.:** Investigation; Data curation; Writing – review and editing. **P.P.:** Conceptualization; Methodology; Supervision; Project administration; Writing – review and editing. R.K.S. and P.K.D.K. independently assigned and reconciled the coverage ratings reported in Tables 5, 8 and 10. All authors read and approved the final manuscript.

### Disclaimer

The affiliations listed reflect the authors' places of employment or academic affiliation for identification purposes only. This research was conducted independently and does not represent the views, opinions, or endorsements of those institutions. No employer resources, proprietary data, or institutional support were used to prepare this manuscript.

## Abbreviations

The following abbreviations are used in this manuscript:

| Abbreviation | Definition |
|---|---|
| A2A | Agent-to-Agent |
| ABAC | Attribute-Based Access Control |
| AI | Artificial Intelligence |
| API | Application Programming Interface |
| DID | Decentralized Identifier |
| IAM | Identity and Access Management |

| Abbreviation | Definition |
|---|---|
| IETF | Internet Engineering Task Force |
| IFC | Information Flow Control |
| JIT | Just-in-Time |
| LLM | Large Language Model |
| MAC | Mandatory Access Control |
| MCP | Model Context Protocol |
| NHI | Non-Human Identity |
| OIDC | OpenID Connect |
| PAP | Policy Administration Point |
| PBAC | Policy-Based Access Control |
| PDP | Policy Decision Point |
| PEP | Policy Enforcement Point |
| PKCE | Proof Key for Code Exchange |
| RAG | Retrieval-Augmented Generation |
| RAR | Rich Authorization Requests |
| RBAC | Role-Based Access Control |
| ReBAC | Relationship-Based Access Control |
| RPA | Robotic Process Automation |
| SaaS | Software as a Service |
| SPIFFE | Secure Production Identity Framework for Everyone |
| VC | Verifiable Credential |
| W3C | World Wide Web Consortium |
| WIMSE | Workload Identity in Multi-System Environments |